\documentclass{article}

\usepackage{graphicx} 
\usepackage{setspace} 
\usepackage{natbib} 
\usepackage{amsmath} 
\usepackage{bm}
\usepackage{float} 
\usepackage{placeins} 
\usepackage{rotating} 
\usepackage{pst-all} 
\usepackage{pdftricks} 
\usepackage{lscape} 
\usepackage{bbm} 
\usepackage{comment} 
\usepackage{authblk} 
\usepackage{array}
\usepackage{longtable} 
\usepackage{caption} 
\usepackage{tikz} 
\usepackage{enumitem} 
\usepackage{subcaption} 
\usepackage{booktabs} 
\usepackage{changepage}
\usepackage{etex}
\usepackage{geometry} 
\usepackage{hyperref} 
\usepackage{lineno} 
\usepackage{xcolor} 
\usepackage{overcite} 
\usepackage{amsmath}
\usepackage{amssymb}
\usepackage[retainorgcmds]{IEEEtrantools}
\usepackage{titling}
\usepackage{multirow}

\hypersetup{
    unicode=false,           
    pdftoolbar=true,         
    pdfmenubar=true,         
    pdffitwindow=true,       
    pdfstartview={FitH},     
    pdftitle={},             
    pdfauthor={},            
    pdfsubject={},           
    pdfcreator={},           
    pdfkeywords={},          
    pdfnewwindow=true,       
    colorlinks=true,         
    linkcolor=black,         
    citecolor=black,         
    filecolor=black,         
    urlcolor=blue            
}

\date{\today}

\begin{document}

\author{
    Jesse Zhou\textsuperscript{1}\thanks{Direct correspondence to: Jesse Zhou, University of Chicago, Department of Sociology, 1126 E. 59th St., Chicago, IL, 60637; jessezhou1@uchicago.edu. This research was supported by the Stone Center for Research on Wealth Inequality and Mobility at the University of Chicago. We used ChatGPT, version 4o, for copyediting, assistance with LaTeX code, and aid in debugging Python and R scripts. Responsibility for all content and any potential errors in this manuscript rests solely with the authors.}  
   \hspace{0.1cm} and Geoffrey T. Wodtke\textsuperscript{1} \vspace{0.75cm}\\
    \textsuperscript{1}\small Department of Sociology, University of Chicago, Chicago, IL, USA
\vspace{0.75cm}}

\begin{titlingpage}

\title{Causal Mediation Analysis with Multiple Mediators: A Simulation Approach \vspace{0.5cm}}

\maketitle

\begin{abstract}

Analyses of causal mediation often involve exposure-induced confounders or, relatedly, multiple mediators. In such applications, researchers aim to estimate a variety of different quantities, including interventional direct and indirect effects, multivariate natural direct and indirect effects, and/or path-specific effects. This study introduces a general approach to estimating all these quantities by simulating potential outcomes from a series of distribution models for each mediator and the outcome. Building on similar methods developed for analyses with only a single mediator (\citealt{imai2010general}), we first outline how to implement this approach with parametric models. The parametric implementation can accommodate linear and nonlinear relationships, both continuous and discrete mediators, and many different types of outcomes. However, it depends on correct specification of each model used to simulate the potential outcomes. To address the risk of misspecification, we also introduce an alternative implementation using a novel class of nonparametric models, which leverage deep neural networks to approximate the relevant distributions without relying on strict assumptions about functional form. We illustrate both methods by reanalyzing the effects of media framing on attitudes toward immigration (\citealt{brader2008triggers}) and the effects of prenatal care on preterm birth (\citealt{vanderweele2014effectdecomp}).

\vspace{0.5cm}


\end{abstract}

\end{titlingpage}

\newpage

\doublespacing

\sloppy 

\section{Introduction}

Causal mediation analysis is central to the social, behavioral, and health sciences (\citealt{hayes2017introduction, mackinnon2008mediationtext, vanderweele2015explanation}; \citeauthor{wodtke20XXmediation} Forthcoming). It aims to decompose a total effect of a treatment on an outcome into indirect effects, transmitted through one or more mediators of interest, and a direct effect, which operates through other pathways. When analyzing multiple mediators, researchers often focus on multivariate natural direct and indirect effects to assess how all mediators collectively explain the treatment's impact on the outcome (\citealt{vanderweele2014effectdecomp, vanderweele2015explanation}); path-specific effects, which disentangle the explanatory role of each mediator in a causal sequence, accounting for the influence of preceding mediators earlier in the causal chain (\citealt{avin2005identifiability, zhou2023tracing, zhou2022semiparametric}); or interventional direct and indirect effects, which isolate the mediating influence of a single mediator when its relationship with the outcome is confounded by a prior mediator, also known as a treatment-induced confounder (\citealt{geneletti2007identifying, didelez2012direct, wodtke2020effect}).

Several methods are currently available for estimating these effects, including procedures based on inverse probability weighting (IPW; \citealt{vanderweele2014effectdecomp}), regression imputation (RI; \citealt{zhou2023tracing}), and linear modeling (\citealt{hayes2017introduction, vanderweele2015explanation, wodtke2020effect}). Each of these approaches, however, suffers from non-trivial limitations. In particular, IPW estimators are challenging to use and are often unstable when the treatment or any of the mediators are not discrete variables with only a few values (\citealt{mortimer2005application, naimi2014constructing, zhou2020residual}). Approaches based on linear modeling are more accommodating of continuous data but are less appropriate when any of the mediators or the outcome are discrete. Moreover, even with continuous data, these methods typically involve arbitrary yet restrictive assumptions about the functional form of the relationships among variables. RI estimators avoid some of these challenges while introducing others. For example, they require multiple different models for the outcome, as well as several auxiliary models for their predictions, which can be difficult to correctly specify in practice. Robust approaches to estimation that combine RI with IPW help mitigate concerns about model misspecification (e.g., \citealt{diaz2021nonparametric, tchetgen2012semiparametric, zhou2022semiparametric}), but these methods are also difficult to implement when the treatment is continuous or otherwise has many values. 

For analyses with only a single mediator, a more general approach---increasingly used in the social sciences---estimates direct and indirect effects by simulating potential outcomes from parametric distribution models for both the mediator and the outcome (\citealt{imai2010general, imai2011unpacking}). This approach offers considerable flexibility, as it accommodates a wide range of models that can easily incorporate linear and nonlinear relationships, interactions, and effect heterogeneity, as well as continuous or discrete variables. However, despite the versatility and broad applicability of the simulation approach, it has yet to be adapted for settings with multiple mediators, where its flexibility could address many of the limitations that hamper the other estimation strategies outlined previously.

In this study, we extend the simulation approach for mediation analyses involving multiple mediators, where the targeted estimands include multivariate natural effects, path-specific effects, and interventional effects. We first introduce simulation estimators for these effects based on parametric distribution models for each mediator and the outcome. These estimators can be implemented with a wide range of parametric models and thus are very flexible. However, their accuracy requires that the models are correctly specified, which can be challenging to achieve in practice, especially when modeling multiple mediators. If any of the distribution models for the mediators or the outcome are misspecified, the resulting estimates may be biased, even when the target estimands are nonparametrically identifiable.

To address this limitation, we also introduce an alternative simulation approach based on a new class of nonparametric models known as normalizing flows (\citealt{wehenkel2019UMNN, wehenkel2020GNF, balgi_deep_2025}). Normalizing flows model the conditional distribution of each mediator and the outcome given their causal antecedents, as specified in a directed acyclic graph (DAG; \citealt{pearl2009}). These models employ a special type of deep neural network that avoids imposing any stringent constraints on the functional form of the conditional distributions or on the relationships among variables. Once trained on data, they can be used to simulate potential outcomes and construct estimates for the effects of interest, similar to the parametric approach.

In the sections that follow, we begin by formally defining a set of target estimands for mediation analyses involving multiple mediators and outline the conditions required for their nonparametric identification. Next, we describe how to estimate these quantities using a simulation approach based on parametric models. We then present a nonparametric approach for constructing simulation estimates using normalizing flows and deep neural networks. To illustrate these methods, we reanalyze whether the effects of media framing on attitudes toward immigration are mediated by anxiety versus perceptions of cost (\citealt{brader2008triggers}), and whether the effects of prenatal care on preterm birth are mediated by maternal smoking and/or pre-eclampsia risk (\citealt{vanderweele2014effectdecomp}). To facilitate adoption, open-source software in R and Python is available at \url{https://github.com/causalMedAnalysis/causalMedR/blob/main/medsim.R} and \url{https://github.com/JesseZhou-1/medflow} for implementing these methods.

\section{Notation, Estimands and Identification} \label{sec:notation}

\begin{figure}[!t]
\begin{centering}
\begin{tikzpicture}[yscale = 2, xscale = 3]

\node[text centered] at (-1,1.25) (c) {$V$};
\node[text centered] at (-1,0) (d) {$D$};
\node[text centered] at (1,0) (x) {$X$};
\node[text centered] at (0,0) (l) {$L$};
\node[text centered] at (2,0) (y) {$Y$};

\draw [->, line width= 1.25] (c) -- (d);
\draw [->, line width= 1.25] (c) -- (l);
\draw [->, line width= 1.25] (c) -- (y);
\draw [->, line width= 1.25] (c) -- (x);
\draw [->, line width= 1.25] (d) -- (l);
\draw [->, line width= 1.25] (d) to [out=-40, in=-140, looseness=1.2] (x);
\draw [->, line width= 1.25] (d) to [out=-40, in=-140, looseness=1.2] (y);
\draw [->, line width= 1.25] (l) -- (x);
\draw [->, line width= 1.25] (l) to [out=-40, in=-140, looseness=1.2] (y);
\draw [->, line width= 1.25] (x) -- (y);

\end{tikzpicture}\caption{A Graphical Mediation Model with Two Mediators. \label{fig:simple_med_model}}
\medskip{}
\par\end{centering}
Note: $D$ denotes the treatment, $L$ denotes an initial mediator, $X$ denotes a second mediator, $Y$ denotes the outcome, and $V$ denotes a set of baseline confounders.
\end{figure}
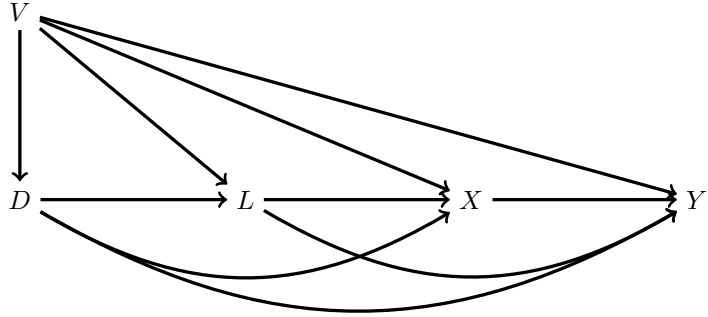

A basic mediation model with two mediators is depicted in Figure \ref{fig:simple_med_model} using a directed acyclic graph (DAG; \citealt{pearl2009, elwert2013graphical}). In a DAG, variables are represented by capital letters, and an arrow from one variable to another denotes a direct causal effect between them. A DAG can be interpreted as a nonparametric structural equation model, which specifies a set of causal relationships among variables without imposing any assumptions about their functional form.

The model in Figure \ref{fig:simple_med_model} includes a treatment $D$, an outcome $Y$, and two mediators $L$ and $X$. The treatment $D$ directly influences both $L$ and $X$, and these mediators in turn directly affect the outcome $Y$. Thus, $L$ and $X$ transmit part of the overall effect of the treatment on the outcome. However, $D$ also has a direct effect on $Y$, indicating that its influence on the outcome is not mediated entirely by $L$ and $X$. Additionally, the first mediator $L$ directly affects both the second mediator $X$ and the outcome $Y$. Because $L$ is influenced by $D$ and affects both $X$ and $Y$, it acts not only as an initial mediator but also as a treatment-induced confounder for the relationship between a second mediator and the outcome. Finally, the model incorporates a set of baseline confounders $V$, which influence all downstream variables.

For simplicity, we focus on a model with two mediators. However, the concepts, estimands, and estimation procedures we describe in this setting are all easily extended for applications involving more than two mediators.

\subsection{Multivariate Natural Direct and Indirect Effects} \label{subsec:mnes}

In analyses with multiple mediators, researchers often aim to decompose the total effect of a treatment on an outcome into two components: one that operates through all the mediators considered jointly and another that does not operate through any of the mediators. This approach parallels the common decomposition of an average total effect into natural direct and indirect effects for a single mediator (\citealt{pearl2001direct, vanderweele2015explanation}), but it generalizes to settings involving a vector of multiple mediators (\citeauthor{wodtke20XXmediation} Forthcoming).

To formally define this multivariate decomposition, we use the potential outcomes framework (\citealt{Holland1986, Rubin1974potential}). Let $d$ represent a specific value of the treatment $D$, and let $Y(d)$ denote the potential value of the outcome if an individual were exposed to treatment $d$, possibly contrary to fact. With this notation, the average total effect of treatment on the outcome can be expressed as follows:
\begin{align}
ATE(d, d^*) &= \mathbb{E}[Y(d) - Y(d^*)], 
\label{eq:ate}
\end{align}
which represents the expected difference in the outcome $Y$ if all individuals were exposed to $d$ instead of an alternative treatment $d^*$.

Now, let \(\mathbf{M} = \{L, X\}\) denote the set of mediators composed of $L$ and $X$ in causal order, and define \(\mathbf{M}(d) = \{L(d), X(d, L(d))\}\) as the potential values of both mediators under exposure to treatment $d$. The vector \(\mathbf{M}(d)\) includes $L(d)$, the potential value of the first mediator under treatment $d$, and $X(d, L(d))$, the potential value of the second mediator under treatment $d$ and, by extension, under $L(d)$ as well. Similarly, let \(Y(d, \mathbf{M}(d)) = Y(d, L(d), X(d,L(d)))\) denote the potential value of the outcome under treatment $d$ and the corresponding levels of both mediators that would naturally arise under this treatment.

Using this notation, the average total effect of treatment on the outcome can be re-expressed and decomposed into direct and indirect components as follows:
\begin{align}
ATE(d, d^*) &= \mathbb{E}[Y(d) - Y(d^*)] \nonumber \\
&= \mathbb{E}[Y(d, \mathbf{M}(d)) - Y(d^*, \mathbf{M}(d^*))] \nonumber \\
&= \underbrace{\mathbb{E}[Y(d, \mathbf{M}(d^*)) - Y(d^*, \mathbf{M}(d^*))]}_{\text{multivariate natural direct effect}} + \underbrace{\mathbb{E}[Y(d, \mathbf{M}(d)) - Y(d, \mathbf{M}(d^*))]}_{\text{multivariate natural indirect effect}}. \label{eq:ate_mne_decomp}
\end{align}
The first term in this decomposition, \(MNDE(d, d^*)=\mathbb{E}[Y(d, \mathbf{M}(d^*)) - Y(d^*, \mathbf{M}(d^*))]\), is the \textit{multivariate natural direct effect}. It captures an effect of treatment \(D\) on the outcome \(Y\) that does not operate through any pathways involving either of the mediators. The second term, \(MNIE(d, d^*)=\mathbb{E}[Y(d, \mathbf{M}(d)) - Y(d, \mathbf{M}(d^*))]\), is the \textit{multivariate natural indirect effect}. This term captures an effect of treatment on the outcome that operates through pathways involving the mediators in \(\mathbf{M}\).

The \(MNDE(d, d^*)\) and \(MNIE(d, d^*)\) can be nonparametrically identified under the following set of assumptions: 
\begin{adjustwidth}{2.5em}{0pt}
    \textbf{MNE.1} --- The potential values of both the outcome and mediators are conditionally independent of the treatment, and the potential outcomes are conditionally independent of the observed mediators and their potential values under the alternative level of treatment. Formally, this assumption can be expressed as $\{Y\left(d, \mathbf{m}\right), \mathbf{M}\left(d\right)\} \perp D|V$, $Y\left(d, \mathbf{m}\right)\perp \mathbf{M}|\{V, D\}$, and $Y\left(d, \mathbf{m}\right)\perp \mathbf{M}\left(d^*\right)|V$ for any $d$, $d^*$, and $\mathbf{m}$, where $Y\left(d, \mathbf{m}\right)$ denotes the potential outcome under treatment $d$ and the values of the mediators given by $\mathbf{m}=\{l,x\}$. In substantive terms, this assumption requires the absence of unobserved confounding for the treatment-mediator, treatment-outcome, and mediator-outcome relationships. Additionally, there must not be any treatment-induced confounding---that is, no confounding of the mediator-outcome relationships by variables that are affected by the treatment, whether they are observed or not. \\
    \textbf{MNE.2} --- There is a positive probability of exposure to all levels of the treatment and mediators, given the baseline confounders. Formally, this condition can be expressed as $P\left(d, \mathbf{m}|v\right) > 0$ for any $d$ and $\mathbf{m}$. \\
    \textbf{MNE.3} --- The observed and potential values of the mediators and outcome are consistent. This condition can be formally expressed as $\mathbf{M}=\mathbf{M}(D)$ and $Y=Y\left(D, \mathbf{M}\right)$. Substantively, it requires that there are not multiple versions of the treatment and mediators with heterogeneous effects on the outcome and that there is no interference between individuals.
\end{adjustwidth}

To identify multivariate natural effects, it suffices to identify the potential outcome means that define them. Let $\phi_{d_1,d_2}=\mathbb{E}[Y(d_2, \mathbf{M}(d_1))$ denote the marginal mean of the potential outcomes under treatment $d_2$ and the values of the mediators that would arise naturally under treatment $d_1$. With this notation, the multivariate natural effects can be expressed as $MNDE(d, d^*)=\phi_{d^*,d}-\phi_{d^*,d^*}$ and $MNIE(d, d^*)=\phi_{d,d}-\phi_{d^*,d}$. When assumptions MNE.1 to MNE.3 are satisfied, $\phi_{d_1,d_2}$ can be identified with the following nonparametric expression:
\begin{align}
\phi_{d_1,d_2} &= \sum_{v,\mathbf{m}} \mathbb{E}\left[Y|v,d_2,\mathbf{m}\right] P\left(\mathbf{m}|v,d_1\right) P\left(v\right) \nonumber \\
&= \sum_{v,l,x} \mathbb{E}\left[Y|v,d_2,l,x\right] P\left(x|v,d_1,l\right) P\left(l|v,d_1\right) P\left(v\right),
\label{eq:phi_np_id_formula}
\end{align}
where $\mathbb{E}\left[Y|v,d_2,l,x\right]$ is the conditional expected value of the observed outcome, $ P\left(\mathbf{m}|v,d_1\right)$ is the joint probability of the mediators conditional on the baseline confounders and the value of treatment given by $d_1$, and $P\left(v\right)$ denotes the marginal probability of the baseline confounders. The second equality in Equation \ref{eq:phi_np_id_formula} follows from the product rule of joint probability.

\subsection{Path-specific Effects}

When the mediators are causally ordered, the total effect of treatment on the outcome can also be decomposed into a series of \textit{path-specific effects} (PSEs; \citealt{vanderweele2014effectdecomp, zhou2023tracing}). Each PSE captures the explanatory role of one mediator, net of the other mediators that precede it in causal order. 

\begin{figure}[!b]
\begin{centering}
\caption{The Causal Paths Connecting Treatment with the Outcome in a Graphical Mediation Model with Two Mediators. \label{fig:med_model_paths}}
\vspace{-1\baselineskip}

\begin{minipage}[t]{0.7\columnwidth}%
\noindent \begin{center}
\begin{tikzpicture}[yscale = 1.5, xscale = 2.5]

\node[text centered] at (-1,1.25) (c) {$V$};
\node[text centered] at (-1,0) (d) {$D$};
\node[text centered] at (1,0) (x) {$X$};
\node[text centered] at (0,0) (l) {$L$};
\node[text centered] at (2,0) (y) {$Y$};

\draw [->, line width= 1.25] (c) -- (d);
\draw [->, line width= 1.25] (c) -- (l);
\draw [->, line width= 1.25] (c) -- (y);
\draw [->, line width= 1.25] (c) -- (x);
\draw [->, line width= 1.25] (d) -- (l);
\draw [->, line width= 1.25] (d) to [out=-40, in=-140, looseness=1.2] (x);
\draw [->, line width= 1.25] (d) to [out=-40, in=-140, looseness=1.2] (y);
\draw [->, line width= 1.25] (l) -- (x);
\draw [->, line width= 1.25] (l) to [out=-40, in=-140, looseness=1.2] (y);
\draw [->, line width= 1.25] (x) -- (y);
\end{tikzpicture}
\par\end{center}
\end{minipage}\vfill{}

\begin{minipage}[t]{0.45\columnwidth}
\noindent \begin{center}
\begin{tikzpicture}[yscale = 1, xscale = 2]

\node[text centered] at (-1.5,0) (p) {(i)};
\node[text centered] at (-1,0) (a) {$D$};
\node[text centered] at (1,0) (m) {$X$};
\node[text centered] at (0,0) (l) {$L$};
\node[text centered] at (2,0) (y) {$Y$};

\draw [->, line width= 1.25] (a) to [out=-30, in=-150, looseness=1.2] (y);
 
\node[text centered] at (-1.5,-2) (p) {(iii)};
\node[text centered] at (-1,-2) (a) {$D$};
\node[text centered] at (1,-2) (m) {$X$};
\node[text centered] at (0,-2) (l) {$L$};
\node[text centered] at (2,-2) (y) {$Y$};

\draw [->, line width= 1.25] (a) -- (l);
\draw [->, line width= 1.25] (l) to [out=-30, in=-150, looseness=1.2] (y);
\end{tikzpicture}
\par\end{center}%

\end{minipage}\qquad{}
\begin{minipage}[t]{0.45\columnwidth}
\noindent \begin{center}
\begin{tikzpicture}[yscale = 1, xscale = 2]

\node[text centered] at (-1.5,0) (p) {(ii)};
\node[text centered] at (-1,0) (a) {$D$};
\node[text centered] at (1,0) (m) {$X$};
\node[text centered] at (0,0) (l) {$L$};
\node[text centered] at (2,0) (y) {$Y$};

\draw [->, line width= 1.25] (a) to [out=-30, in=-150, looseness=1.2] (m);
\draw [->, line width= 1.25] (m) -- (y);

\node[text centered] at (-1.5,-2) (p) {(iv)};
\node[text centered] at (-1,-2) (a) {$D$};
\node[text centered] at (1,-2) (m) {$X$};
\node[text centered] at (0,-2) (l) {$L$};
\node[text centered] at (2,-2) (y) {$Y$};

\draw [->, line width= 1.25] (a) -- (l);
\draw [->, line width= 1.25] (l) -- (m);
\draw [->, line width= 1.25] (m) -- (y);
\end{tikzpicture}
\par\end{center}
\end{minipage}
\par\end{centering}
\medskip{}
Note: $D$ denotes the treatment, $L$ denotes an initial mediator, $X$ denotes a second mediator, $Y$ denotes the outcome, and $V$ denotes a set of baseline confounders. The confounding paths between $V$ and the other variables are omitted in subgraphs (i) to (iv).
\end{figure}
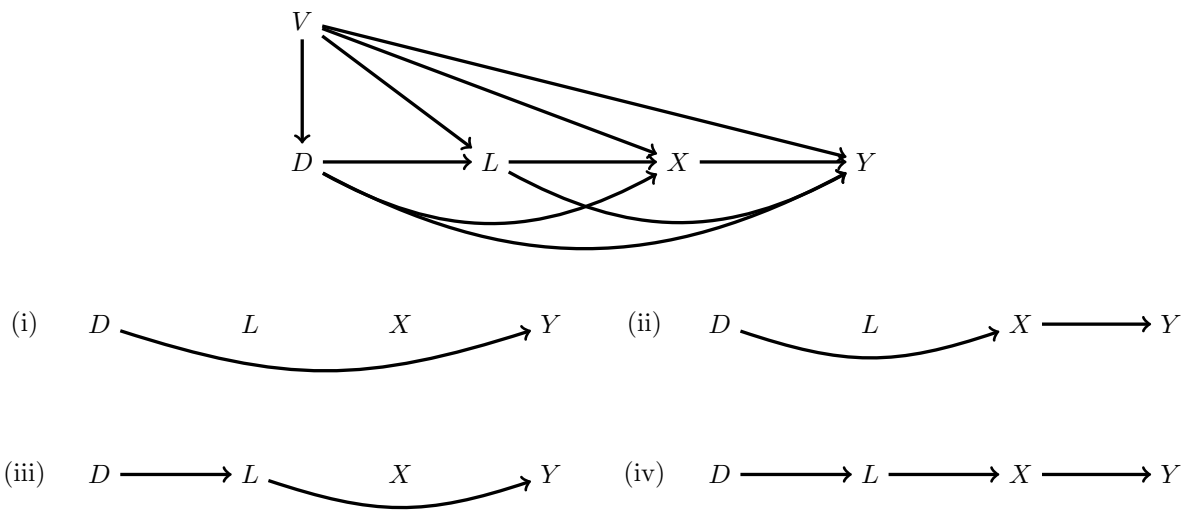

In our graphical model from the previous section, there are four causal paths from treatment $D$ to the outcome $Y$: (i) \(D \rightarrow Y\), the effect not mediated by either $L$ or $X$; (ii) \(D \rightarrow X \rightarrow Y\), the effect mediated only by $X$; (iii) \(D \rightarrow L \rightarrow Y\), the effect mediated only by $L$; and (iv) \(D \rightarrow L \rightarrow X \rightarrow Y\), the effect mediated sequentially through $L$ and $X$. These paths are illustrated in Figure \ref{fig:med_model_paths}, which reproduces the structure of our original DAG while highlighting each of the causal paths connecting treatment with the outcome. 

In this setting, where two causally ordered mediators transmit the effect of treatment on the outcome, the $ATE(d,d^*)$ can be decomposed into a series of PSEs as follows:
\begin{align}
ATE(d, d^*) &= \mathbb{E}[Y(d) - Y(d^*)] \nonumber \\
&= \mathbb{E}[Y(d, L(d), X(d, L(d))) - Y(d^*, L(d^*), X(d^*, L(d^*)))] \nonumber \\
&= \underbrace{\mathbb{E}[Y(d, L(d^*), X(d^*, L(d^*))) - Y(d^*, L(d^*), X(d^*, L(d^*)))]}_{D \rightarrow Y} \nonumber \\
&\qquad + \underbrace{\mathbb{E}[Y(d, L(d^*), X(d, L(d^*))) - Y(d, L(d^*), X(d^*, L(d^*)))]}_{D \rightarrow X \rightarrow Y} \nonumber \\
&\qquad + \underbrace{\mathbb{E}[Y(d, L(d), X(d, L(d))) - Y(d, L(d^*), X(d, L(d^*)))]}_{D \rightarrow L \rightarrow Y; D \rightarrow L \rightarrow X \rightarrow Y} \nonumber \\
&= PSE_{D \rightarrow Y}(d, d^*) + PSE_{D \rightarrow X \rightarrow Y}(d, d^*) + PSE_{D \rightarrow L \rightsquigarrow Y}(d, d^*). 
\label{eq:pse_decomp}
\end{align}
The three terms in this decomposition---\(PSE_{D \rightarrow Y}(d, d^*)\), \(PSE_{D \rightarrow X \rightarrow Y}(d, d^*)\), and \(PSE_{D \rightarrow L \rightsquigarrow Y}(d, d^*)\)---correspond to the effects of treatment transmitted along distinct causal paths, with straight arrows denoting a single causal path and squiggly arrows representing a combination of multiple paths.

Specifically, the first term, \(PSE_{D \rightarrow Y}(d, d^*)\), represents the direct effect of treatment on the outcome (\(D \rightarrow Y\)). This PSE is equivalent to the $MNDE(d, d^*)$, as it captures an effect of treatment on the outcome that is not mediated by either $L$ or $X$. The second term, \(PSE_{D \rightarrow X \rightarrow Y}(d, d^*)\), captures the effect mediated exclusively through \(X\) (\(D \rightarrow X \rightarrow Y\)). The third term, \(PSE_{D \rightarrow L \rightsquigarrow Y}(d, d^*)\), represents the effect of treatment transmitted through $L$, including both the component mediated solely by $L$ (\(D \rightarrow L \rightarrow Y\)) and the component mediated sequentially through $L$ and $X$ (\(D \rightarrow L \rightarrow X \rightarrow Y\)). In other words, the \(PSE_{D \rightarrow L \rightsquigarrow Y}(d, d^*)\) captures the \textit{overall} mediating effect of \(L\), part of which may also operate through \(X\). The sum of \(PSE_{D \rightarrow X \rightarrow Y}(d, d^*)\) and \(PSE_{D \rightarrow L \rightsquigarrow Y}(d, d^*)\) is equal to the $MNIE(d, d^*)$, which captures an effect of treatment on the outcome mediated jointly by $L$ and $X$ together.

Although four causal paths connect \(D\) to \(Y\) in our graphical model, Equation \ref{eq:pse_decomp} partitions the total effect into only three components. The last component, \(PSE_{D \rightarrow L \rightsquigarrow Y}(d, d^*)\), encompasses two different causal paths: \(D \rightarrow L \rightarrow Y\) and \(D \rightarrow L \rightarrow X \rightarrow Y\). If the mediators \(L\) and \(X\) were causally independent, the second path, \(D \rightarrow L \rightarrow X \rightarrow Y\), would not exist. In this case, the total effect of \(D\) on \(Y\) could be partitioned into an effect mediated through \(L\) alone (\(D \rightarrow L \rightarrow Y\)), an effect mediated through \(X\) alone (\(D \rightarrow X \rightarrow Y\)), and a direct effect operating through neither \(L\) nor \(X\) (\(D \rightarrow Y\)). However, in the general case where \(L\) and \(X\) are causally dependent, it is not possible to separate the mediating effects of \(L\) and \(X\) into unique components, as the total effect is partly transmitted by \(L\) and \(X\) operating jointly in a causal chain (\citealt{zhou2023tracing}).

The PSEs defined previously can be nonparametrically identified if the following conditions are met: 
\begin{adjustwidth}{2.5em}{0pt}
    \textbf{PSE.1} --- For each set of mediators in causal order, the potential values of both the outcome and mediators must be conditionally independent of the treatment. In addition, the potential outcomes must also be conditionally independent of the observed mediators and their potential values under the alternative level of treatment. Formally, this assumption can be expressed as $\{Y\left(d, l\right), Y\left(d, l, x\right), L\left(d\right), X\left(d\right)\} \perp D|V$; $Y\left(d, l\right) \perp L|\{V, D\}$; $Y\left(d, l, x\right) \perp \{L,X\}|\{V, D\}$; $Y\left(d,l\right) \perp L\left(d^*\right)|V$; and $Y\left(d,l,x\right) \perp \{L\left(d^*\right), X\left(d^*,L(d^*)\right) \}|V$ for any $d$, $d^*$, $l$, and $x$. In substantive terms, this assumption requires the absence of unobserved confounding for the treatment-mediator, treatment-outcome, mediator-outcome, and mediator-mediator relationships. It also requires that there must not be any treatment-induced confounding of the mediator-outcome or mediator-mediator relationships, whether observed or not. \\
    \textbf{PSE.2} --- There must be a positive probability for all values of the exposure given the baseline confounders and each set of mediators in causal order. Formally, this condition can be expressed as $P\left(d|v\right) > 0$, $P\left(d|v,l\right)>0$ and $P\left(d|v,l,x\right)>0$ for any $d$. \\
    \textbf{PSE.3} --- The observed and potential values of the mediators and outcome must be consistent, which requires that $L=L(D)$, $X=X(D,L)$, and $Y=Y\left(D,L,X\right)$.
\end{adjustwidth}

To identify PSEs, it is sufficient to identify the potential outcome means that define them. Let $\psi_{d_1, d_2, d_3} = \mathbb{E}[Y(d_3, L(d_1), X(d_2, L(d_1)))]$ represent the marginal mean of the potential outcomes under treatment $d_3$, the value of the first mediator $L$ that would arise naturally under treatment $d_1$, and the value of the second mediator $X$ that would arise naturally under treatment $d_2$ and $L(d_1)$. With this notation, the PSEs in Equation \ref{eq:pse_decomp} can be expressed as follows:
\begin{align}
ATE(d, d^*) 
&= \underbrace{\psi_{d^*, d^*, d} - \psi_{d^*, d^*, d^*}}_{PSE_{D \rightarrow Y}(d, d^*)} + \underbrace{\psi_{d^*, d, d} - \psi_{d^*, d^*, d}}_{PSE_{D \rightarrow X \rightarrow Y}(d, d^*)} + \underbrace{\psi_{d, d, d} - \psi_{d^*, d, d}}_{PSE_{D \rightarrow L \rightsquigarrow Y}(d, d^*)}. \label{eq:pse_decomp_psi}
\end{align}
When assumptions PSE.1 to PSE.3 are satisfied, $\psi_{d_1,d_2,d_3}$ can be identified using the following nonparametric expression (\citealp{avin2005identifiability, zhou2022semiparametric}):
\begin{align}
\psi_{d_1, d_2, d_3} 
&= \sum_{v,l,x}\mathbb{E}[Y|v, d_3, l, x] P(x|v, d_2, l) P(l|v, d_1) P(v). \label{eq:psi_np_id_formula}
\end{align}
In this identification formula, $\mathbb{E}\left[Y|v,d_3,l,x\right]$ is the conditional expected value of the observed outcome given $\{v,d_3,l,x\}$, $P\left(x|v,d_2,l\right)$ is the conditional probability of the second mediator given $\{v,d_2,l\}$, $P\left(l|v,d_1\right)$ is the conditional probability of the first mediator given $\{v,d_1\}$, and $P\left(v\right)$ denotes the marginal probability of the baseline confounders, as before.

\subsection{Interventional Direct and Indirect Effects} \label{subsec:ides}

In some cases, researchers wish to isolate the explanatory role of a single mediator, when other potentially confounding mediators may be present. Consider again the graphical model shown in Figure \ref{fig:simple_med_model}. In this model, analyzing how the treatment $D$ affects the outcome $Y$ through the mediator $X$ is complicated by the presence of the prior mediator $L$, which acts as a treatment-induced confounder for the relationship between $X$ and $Y$. Specifically, when mediators like $L$ function as treatment-induced confounders for a downstream mediator like $X$, the univariate natural direct and indirect effects of treatment operating through $X$ cannot be nonparametrically identified, even when these confounders are observed and can be controlled (\citealt{vanderweele2014effectdecomp, wodtke2020effect}).

In response to these challenges, researchers may instead target a distinct set of estimands known as interventional direct and indirect effects (\citealt{didelez2012direct, geneletti2007identifying, wodtke2020effect}), which can still be nonparametrically identified and consistently estimated when treatment-induced confounders are present. These estimands circumvent the challenges that stem from treatment-induced confounders by focusing on counterfactuals that set the focal mediator to values randomly drawn from its distribution under a specific treatment.

To elaborate, let $\mathcal{X}(d|V)$ represent a value of the mediator $X$ randomly selected from its distribution under treatment $d$ among individuals with baseline confounders $V$. Here, the script $\mathcal{X}$ is used to distinguish this randomly selected value from $X(d)=X(d,L(d))$, the potential value of the mediator that would naturally arise for a given individual following exposure to treatment $d$. By extension, let $Y(d, \mathcal{X}(d|V))$ denote the potential value of the outcome that would occur if an individual were exposed to treatment $d$ and then had experienced a level of the mediator $X$ randomly selected from its distribution under exposure $d$. We refer to quantities like $Y(d, \mathcal{X}(d|V))$ as randomized potential outcomes because they involve random draws from a distribution of the focal mediator (\citeauthor{wodtke20XXmediation} Forthcoming). 

Interventional direct and indirect effects are defined in terms of contrasts between randomized potential outcomes. Specifically, the \textit{interventional direct effect} of treatment $D$ on the outcome $Y$ is formally defined as follows:
\begin{equation}
IDE\left(d,d^{*}\right)=\mathbb{E}\left[Y\left(d,\mathcal{X}\left(d^{*}|V\right)\right)-Y\left(d^{*},\mathcal{X}\left(d^{*}|V\right)\right)\right]. \label{eq:ide}
\end{equation}
This quantity represents the expected difference in the outcome if individuals were exposed to treatment $d$ rather than $d^{*}$, and had then experienced a level of the mediator that was randomly selected from its distribution under treatment $d$. The $IDE\left(d,d^{*}\right)$ captures an effect of treatment on the outcome that does not operate through the mediator of interest $X$.

Similarly, the \textit{interventional indirect effect} is formally defined as follows:
\begin{equation}
IIE\left(d,d^{*}\right)=\mathbb{E}\left[Y\left(d,\mathcal{X}\left(d|V\right)\right)-Y\left(d,\mathcal{X}\left(d^{*}|V\right)\right)\right].\label{eq:iie}
\end{equation}
This quantity represents the expected difference in the outcome if individuals were exposed to treatment $d$, and had then experienced a level of the mediator randomly selected from its distribution under treatment $d$ rather than its distribution under treatment $d^{*}$. The $IIE\left(d,d^{*}\right)$ captures an effect of treatment on the outcome transmitted through $X$.

The interventional direct and indirect effects sum to an analogue of the average total effect, defined in terms of the randomized potential outcomes. This analogue, known as the \textit{overall effect} of treatment $D$ on the outcome $Y$ (\citealt{nguyen2022clarifying,vanderweele2014effectdecomp}), is formally expressed as follows:
\begin{align}
OE\left(d,d^{*}\right) & =IDE\left(d,d^{*}\right)+IIE\left(d,d^{*}\right)\nonumber \\
 & =\mathbb{E}\left[Y\left(d,\mathcal{X}\left(d^{*}|V\right)\right)-Y\left(d^{*},\mathcal{X}\left(d^{*}|V\right)\right)\right]+\mathbb{E}\left[Y\left(d,\mathcal{M}\left(d|V\right)\right)-Y\left(d,\mathcal{X}\left(d^{*}|V\right)\right)\right]\nonumber \\
 & =\mathbb{E}\left[Y\left(d,\mathcal{X}\left(d|V\right)\right)-Y\left(d^{*},\mathcal{X}\left(d^{*}|V\right)\right)\right].\label{eq:oe}
\end{align}
The $OE\left(d,d^{*}\right)$ contrasts randomized potential outcomes under different levels of both the treatment $D$ and the mediator $X$, where values for $X$ are drawn from its distribution corresponding to each of the treatments compared. This quantity captures the combined impact of a change in treatment and the resulting shift in the distribution of the focal mediator at the population level.

Nonparametric identification of the $IDE\left(d,d^{*}\right)$,  $IIE\left(d,d^{*}\right)$, and $OE\left(d,d^{*}\right)$ can be achieved under the following set of assumptions:
\begin{adjustwidth}{2.5em}{0pt}
    \textbf{(IE.1)} --- The potential values of both the outcome and the focal mediator are conditionally independent of the treatment, and the potential outcomes are conditionally independent of the observed mediator. Formally, this assumption can be expressed as $\{Y\left(d, x\right), X\left(d\right)\} \perp D|V$ and $Y\left(d, x\right)\perp X|V, D, L$ for any $d$ and $x$. In substantive terms, it requires that there must not be any unobserved confounding for the treatment-mediator, treatment-outcome, and mediator-outcome relationships. \\
    \textbf{(IE.2)} --- The probability of exposure to all levels of the treatment and the focal mediator must be sequentially positive. Formally, this condition can be expressed as $P\left(d|v\right) > 0$ and $P\left(x|v, d, l\right) > 0$ for any $d$ and $x$. \\
    \textbf{(IE.3)} --- The observed and potential values of the focal mediator and the outcome are consistent. This condition can be formally expressed as $X = X\left(D\right)$ and $Y = Y\left(D, X\right)$.
\end{adjustwidth}

To identify interventional effects, it is sufficient to identify the potential outcome means that define them. Let $\lambda_{d_1,d_2}=\mathbb{E}[Y(d_2, \mathcal{X}\left(d_1|V\right)]$ denote the marginal mean of the randomized potential outcomes under treatment $d_2$ and a value of the mediator $X$ randomly drawn from its distribution under treatment $d_1$. Using this notation, the interventional direct and indirect effects can be expressed as $IDE(d, d^*)=\lambda_{d^*,d}-\lambda_{d^*,d^*}$ and $IIE(d, d^*)=\lambda_{d,d}-\lambda_{d^*,d}$, respectively. When assumptions IE.1 to IE.3 hold, $\lambda_{d_1,d_2}$ can be nonparametrically identified with the following expression:
\begin{equation}
\lambda_{d_1,d_2}=\sum_{v,x,l}\mathbb{E}\left[Y|v,d_2,l,x\right]P\left(l|v,d_2\right) P\left(x|v,d_1\right)P\left(v\right), \label{eq:lambda_np_id_formula}
\end{equation}
where $P\left(x|v,d_1\right)$ represents the conditional probability that $X=x$ among individuals with $V=v$ and $D=d_1$, and all other terms are defined as before.

\section{Estimation with Parametric Models\label{sec:Parametric-Estimation}}

In this section, we describe a simulation approach for estimating interventional, path-specific, and multivariate natural effects. The procedure begins by fitting parametric distribution models to the observed data for each mediator and the outcome. The fitted models are then used to simulate values for these variables under different conditions using a Monte Carlo sampling procedure. By averaging these simulated values over many replications, we approximate the complex, probability-weighted sums that constitute the identification formulas outlined previously.

This approach is compatible with a wide range of parametric models for the mediators and the outcome, including nonlinear and non-additive specifications that render direct evaluation of the identification formulas intractable. The resulting estimates are consistent, meaning they converge to their target estimands as both the sample size and the number of simulations increase toward infinity, provided that all models used for generating the simulated values are correctly specified. Confidence intervals and hypothesis tests can be constructed using the nonparametric bootstrap.

Specifically, for estimating multivariate natural and path-specific effects, the simulation approach is implemented through the following steps:
\begin{enumerate}
\item \textbf{Fit models for each mediator and the outcome.} Begin by fitting a parametric distribution model for the first mediator conditional on the baseline confounders and the treatment, denoted by \(g_{L|V,D}(V, D)\). Next, fit a distribution model for the second mediator, conditional on the baseline confounders, treatment, and the first mediator, denoted by \(g_{X|V,D,L}(V, D, L)\). Then, fit a distribution model for the outcome, conditional on the baseline confounders, treatment, and both mediators, denoted by \(g_{Y|V,D,L,X}(V, D, L, X)\). Let \(\hat{g}_{L|V,D}(V, D)\), \(\hat{g}_{X|V,D,L}(V, D, L)\), and \(\hat{g}_{Y|V,D,L,X}(V, D, L, X)\) represent these models with their parameters estimated by maximum likelihood.

\item \textbf{Simulate potential values for the first mediator.} For each individual in the sample, simulate \(J\) copies of \(L(d_1)\) from \(\hat{g}_{L|V,D}(V, d_1)\). This is accomplished by replacing each sample member's treatment with the value \(d_1\), while leaving their baseline confounders, \(V\), unchanged. Monte Carlo samples are then drawn from the fitted model with the predictors set at these values. Let \(\tilde{L}_{j}(d_1)\) denote the simulated values of the first mediator for each simulation \(j = 1, 2, \dots, J\).

\item \textbf{Simulate potential values for the second mediator.} For each sample member and each simulated value of the first mediator, simulate one copy of \(X(d_2, L(d_1))\) from \(\hat{g}_{X|V,D,L}(V, d_2, \tilde{L}_{j}(d_1))\). This is accomplished by replacing each sample member’s treatment with the value \(d_2\), replacing the first mediator with a simulated value \(\tilde{L}_{j}(d_1)\), and keeping the baseline confounders \(V\) unchanged. Monte Carlo samples are then drawn from the fitted model with the predictors set at these values. Let \(\tilde{X}_{j}(d_2, L(d_1))\) denote the simulated values of the second mediator for each simulation \(j = 1, 2, \dots, J\).

\item \textbf{Simulate potential outcomes.} For each sample member and each set of simulated mediators, simulate one copy of \(Y(d_3, L(d_1), X(d_2, L(d_1)))\) from \(\hat{g}_{Y|V,D,L,X}(V, d_3, \tilde{L}_{j}(d_1), \tilde{X}_{j}(d_2, L(d_1)))\). This is accomplished by replacing each sample member’s treatment with the value \(d_3\), replacing their mediators with the simulated values \(\tilde{L}_{j}(d_1)\) and \(\tilde{X}_{j}(d_2, L(d_1))\), and leaving the baseline confounders \(V\) unchanged. Monte Carlo samples of the outcome are then drawn from the fitted model with the predictors set at these values. Let \(\tilde{Y}_{j}(d_3, L(d_1), X(d_2, L(d_1)))\) denote the simulated values of the outcome for each simulation \(j = 1, 2, \dots, J\).

\item \textbf{Estimate the marginal mean of the potential outcomes.} Average the simulated outcomes across both simulations and sample members as follows:
\begin{align}
\hat{\psi}_{d_1,d_2,d_3} 
&= \hat{\mathbb{E}}\left[Y(d_3, L(d_1), X(d_2, L(d_1)))\right] \nonumber \\
&= \frac{1}{nJ} \sum \sum_{j} \tilde{Y}_{j}(d_3, L(d_1), X(d_2, L(d_1))),
\label{eq:psi_hat}
\end{align}
where the inner sum is over the \(J\) simulations and the outer sum is over the \(n\) sample members.
\end{enumerate}
By repeating steps 2 to 5 with different values for $d_1$, $d_2$, and $d_3$, we can then contrast the resulting means to obtain estimates for the path-specific effects of interest, which are given by the following expressions:
\begin{align}
\widehat{PSE}_{D \rightarrow Y}(d, d^*) &= \hat{\psi}_{d^*, d^*, d} - \hat{\psi}_{d^*, d^*, d^*} \nonumber \\
\widehat{PSE}_{D \rightarrow X \rightarrow Y}(d, d^*) &= \hat{\psi}_{d^*, d, d} - \hat{\psi}_{d^*, d^*, d} \nonumber \\
\widehat{PSE}_{D \rightarrow L \rightsquigarrow Y}(d, d^*) &= \hat{\psi}_{d, d, d} - \hat{\psi}_{d^*, d, d}.
\label{eq:pse_hat}
\end{align}
By extension, estimates for the multivariate natural effects can be computed as follows: 
\begin{align}
\widehat{MNDE}(d,d^*) &= \widehat{PSE}_{D \rightarrow Y}(d, d^*) \nonumber \\
\widehat{MNIE}(d,d^*) &= \widehat{PSE}_{D \rightarrow X \rightarrow Y}(d, d^*) + \widehat{PSE}_{D \rightarrow L \rightsquigarrow Y}(d, d^*),
\label{eq:mie_hat}
\end{align}
and an estimate for the total effect is given by their sum, $\widehat{ATE}(d,d^*)=\widehat{MNDE}(d,d^*)+\widehat{MNIE}(d,d^*)$.

To estimate interventional effects through the second mediator, while adjusting for treatment-induced confounding by the first mediator, this procedure is modified slightly and implemented as follows (\citeauthor{wodtke20XXmediation} Forthcoming):
\begin{enumerate}[label=\arabic*{*}.]
\item \textbf{Fit models for the treatment-induced confounder, focal mediator, and outcome.} Begin by fitting a parametric distribution model for the treatment-induced confounder, conditional on the baseline confounders and the treatment, denoted by \(g_{L|V,D}(V, D)\). Next, fit a distribution model for the focal mediator conditional on the baseline confounders and treatment, denoted by \(g_{X|V,D}(V, D)\). Then, fit a distribution model for the outcome, conditional on the baseline confounders, the treatment-induced confounder, and the focal mediator, denoted by \(g_{Y|V,D,L,X}(V, D, L, X)\). Let \(\hat{g}_{L|V,D}(V, D)\), \(\hat{g}_{X|V,D}(V, D)\), and \(\hat{g}_{Y|V,D,L,X}(V, D, L, X)\) represent these models with their parameters estimated by maximum likelihood.

\item \textbf{Simulate potential values for the treatment-induced confounder.} For each individual in the sample, simulate \(J\) copies of \(L(d_2)\) from \(\hat{g}_{L|V,D}(V, d_2)\). This is accomplished by replacing each sample member's treatment with the value \(d_2\), while leaving their baseline confounders, \(V\), unchanged. Monte Carlo samples are then drawn from the fitted model with the predictors set at these values. Let \(\tilde{L}_{j}(d_2)\) denote the simulated values of the treatment-induced confounder for each simulation \(j = 1, 2, \dots, J\).

\item \textbf{Simulate potential values for the focal mediator.} For each sample member, simulate \(J\) copies of \(\mathcal{X}\left(d_1|V\right)\) from \(\hat{g}_{X|V,D}(V, d_1)\). This is accomplished by replacing each sample member’s treatment with the value \(d_1\), while leaving their baseline confounders, \(V\), unchanged. Monte Carlo samples are then drawn from the fitted model with the predictors set at these values. Let \(\tilde{\mathcal{X}}_j\left(d_1|V\right)\) denote the simulated values of the second mediator for each simulation \(j = 1, 2, \dots, J\).

\item \textbf{Simulate potential outcomes.} For each sample member and each set of \(\{\tilde{L}_{j}(d_2),\tilde{\mathcal{X}}_j\left(d_1|V\right)\}\), simulate one copy of \(Y(d_2, \mathcal{X}\left(d_1|V\right))\) from \(\hat{g}_{Y|V,D,L,X}(V, d_2, \tilde{L}_{j}(d_2), \tilde{\mathcal{X}}_j\left(d_1|V\right))\). This is accomplished by replacing each sample member’s treatment with the value \(d_2\), replacing their mediators with the simulated values \(\tilde{L}_{j}(d_2)\) and \(\tilde{\mathcal{X}}_j\left(d_1|V\right)\), and leaving the baseline confounders, \(V\), unchanged. Monte Carlo samples of the outcome are then drawn from the fitted model with the predictors set at these values. Let \(\tilde{Y}_{j}(d_2, \mathcal{X}\left(d_1|V\right))\) denote the simulated values of the outcome for each simulation \(j = 1, 2, \dots, J\).

\item \textbf{Estimate the marginal mean of the potential outcomes.} Average the simulated outcomes across both simulations and sample members as follows:
\begin{align}
\hat{\lambda}_{d_1,d_2} 
&= \hat{\mathbb{E}}\left[Y(d_2, \mathcal{X}\left(d_1|V\right))\right] \nonumber \\
&= \frac{1}{nJ} \sum \sum_{j} \tilde{Y}_{j}(d_2, \mathcal{X}\left(d_1|V\right)),
\label{eq:lambda_hat}
\end{align}
where the inner sum is over the \(J\) simulations and the outer sum is over the \(n\) sample members.
\end{enumerate}
Repeating steps 2* to 5* with different values for $d_1$ and $d_2$ yields a set of marginal means that can then be compared to obtain estimates for the interventional effects of interest. Specifically, estimates for the interventional direct and indirect effects are given by the following expressions:
\begin{align}
\widehat{IDE}(d, d^*) &= \hat{\lambda}_{d^*,d} - \hat{\lambda}_{d^*, d^*} \nonumber \\
\widehat{IEE}(d, d^*) &= \hat{\lambda}_{d, d} - \hat{\lambda}_{d^*, d}, 
\label{eq:ie_hat}
\end{align}
and an estimate for the overall effect is given by their sum, $\widehat{OE}(d, d^*)=\widehat{IDE}(d, d^*)+\widehat{IIE}(d, d^*)$.

A key distinction between the estimation procedures for multivariate natural and path-specific effects, and those for interventional effects, lies in the first step. In step 1 of the procedure for multivariate and path-specific effects, a model is fit for the distribution of the second mediator, $X$, conditional on the baseline confounders $V$, treatment $D$, and the first mediator $L$. Then, when simulating values for the mediators in subsequent steps, the simulated values of $L$ are used to generate those for $X$. As a result, these simulated values are not independent of each other, conditional on the baseline confounders and treatment.

In contrast, step 1* of the procedure for estimating interventional effects involves fitting a model for the distribution of the second mediator, $X$, conditional only on the baseline confounders $V$ and treatment $D$. The prior mediator, $L$, is now considered a treatment-induced confounder and excluded from this model. Consequently, when drawing Monte Carlo samples in subsequent steps, the values of $L$ and $X$ are generated independently of one another, conditional on the baseline confounders and treatment. 

This distinction between the two estimation procedures reflects the form of the identification formulas for their respective estimands, which are approximated using averages of Monte Carlo samples drawn from a corresponding set of distribution models. For example, in the identification formula for path-specific effects (Equation \ref{eq:psi_np_id_formula}), the conditional mean of the outcome is averaged over the distributions $P(x|v,d_2,l)$ and $P(l|v,d_1)$. By contrast, the identification formula for interventional effects (Equation \ref{eq:lambda_np_id_formula}) averages the conditional mean of the outcome over $P(x|v,d_1)$ and $P(l|v,d_2)$, where the distribution of $X$ is expressed without conditioning on $L$.

These estimation procedures are highly versatile and can accommodate a wide range of distribution models for the mediators and outcome. For example, if $L$ is binary, $X$ is continuous, and $Y$ is a count, suitable choices might include a logistic regression model for $L$, a normal linear model for $X$, and a Poisson model for $Y$. The linear predictors in these models can also incorporate interaction terms, polynomials, transformations, and/or splines to capture more complex relationships among the variables. To ensure valid inferences about mediation, the models must be correctly specified, with finite-dimensional parameters that can be consistently estimated via maximum likelihood. The only other requirement is that these models must be generative, to enable Monte Carlo sampling. In practice, we recommend using at least $J=10^3$ Monte Carlo samples, with larger values offering greater precision at the cost of additional computation time.

\section{Estimation with Deep Neural Networks}

While the parametric approach to estimation offers significant flexibility, its accuracy depends on correctly specifying parametric models for each mediator and the outcome. Correctly specifying these models can be challenging in practice, particularly in applications with many baseline confounders and more than two mediators. If any of the distribution models for the mediators or the outcome are misspecified, the resulting estimates may be biased, even when the target estimands are nonparametrically identified.

In this section, we introduce an alternative approach to estimation that closely follows the steps outlined previously, but replaces the parametric models with a specific class of deep neural networks (\citealt{balgi_deep_2025, wehenkel2019UMNN, wehenkel2020GNF}). In general, deep neural networks consist of interconnected nodes organized into layers: an input layer, one or more ``hidden" layers, and an output layer. Nodes in each hidden layer receive inputs from the preceding layer, process them using a set of weights and an activation function, and then produce outputs, which serve as inputs for the next layer. The network’s architecture is determined by the number of layers, the number of nodes in each layer, and the weights and activation functions governing the connections between them. With sufficiently flexible architectures, deep neural networks are universal function approximators, capable of modeling any continuous function to an arbitrary degree of accuracy (\citealt{hornik1989universal}).

The deep neural networks we use here model the conditional distributions of each mediator and the outcome, given their causal antecedents, without imposing strict assumptions on the functional form of these distributions or the relationships among variables. After training on the sample data, they can be used to simulate potential values for the mediators and the outcome, following the same procedures described earlier. Thus, the key distinction between these two approaches to estimation lies in the choice of models: the parametric approach relies on distribution models with a prescribed functional form, whereas the approach based on neural networks employs flexible, data-adaptive models that are capable of approximating almost any probability distribution with high accuracy.

Specifically, our alternative approach to estimation uses a special type of \textit{normalizing flow} (\citealt{balgi_deep_2025, papamakarios2021NF_pmi, wehenkel2020GNF}) to model the distributions of the mediators and the outcome. This approach leverages an unconstrained monotonic neural network (UMNN; \citealt{Huang2018NAF, wehenkel2019UMNN}) to transform variables with potentially complex probability distributions into variables with simpler, standard normal distributions. The normalizing transformation is designed so that it is both differentiable and invertible, which enables the computation of probability densities through the change-of-variables formula and allows for efficient generation of Monte Carlo samples by passing standard normal draws through the inverse of the transformation.

To illustrate, consider modeling the distribution of the first mediator $L$ conditional on the baseline confounders $V$ and the treatment $D$. In this context, we define a normalizing flow for $L$ as follows:
\begin{align}
Z_L &= h_{L|V,D}(V, D, L) \sim \mathcal{N}(0,1),
\label{eq:norm_flow_for_L}
\end{align}
where $h_{L|V,D}(V, D, L)$ is a transformation that maps $L$ to a new variable $Z_L$ that follows a standard normal distribution, conditional on $V$ and $D$. In words, this transformation shifts the distribution of $L$ as a function of $V$ and $D$ while simultaneously adjusting its shape to conform to a standard normal curve.

In general, the function $h_{L|V,D}(\cdot)$ may take any form provided it is invertible, and both the transformation and its inverse are smooth with finite first derivatives. This requirement guarantees that each value of $L$ is uniquely mapped to a corresponding value of $Z_L$. Moreover, the invertibility of $h_{L|V,D}(\cdot)$ ensures that $L$ can be recovered from $Z_L$ via the inverse of the normalizing transformation:
\begin{align}
L &= h_{L|V,D}^{-1}(V, D, Z_L) \sim f_{L|V,D},
\label{eq:inv_norm_flow_for_L}
\end{align}
where $f_{L|V,D}$ denotes the true but unknown conditional distribution of $L$ given $V$ and $D$.

When $L$ is discrete rather than continuous, a normalizing flow can still be constructed by employing dequantization (\citealt{balgi2022cgnf, Uria2013RNADETR_uniformdeq, ZieglerR2019VAE_dequant}). With this approach, integer values of a discrete variable are transformed into continuous values by adding a small amount of random noise, typically drawn from a uniform or normal distribution. For example, normal dequantization transforms a discrete variable into a continuous one that follows a multi-modal mixture of normal distributions, with sharply peaked modes at each of the original discrete values. This dequantized variable can then be mapped to a standard normal variable using a normalizing flow. Conversely, the dequantized variable can be recovered from the standard normal variable by applying the inverse of the flow in turn. And because the added noise is minuscule, rounding the dequantized variable to its nearest integer restores the original discrete values almost exactly. In this way, dequantization allows normalizing flows to accurately model a wide range of discrete data, including binary, polytomous, and count variables.

Normalizing flows are generative models. The invertibility of the function $h_{L|V,D}(\cdot)$ enables straightforward and efficient Monte Carlo sampling from the target distribution $f_{L|V,D}$. Specifically, to simulate values from $f_{L|V,D}$, we simply draw Monte Carlo samples from a standard normal distribution and then apply the inverse transformation $h_{L|V,D}^{-1}(\cdot)$ to them.

A normalizing flow also encodes the conditional distribution of $L$ through the change-of-variables formula. In particular, the conditional distribution of $L$ given $V$ and $D$ can be expressed as follows:
\begin{align}
f_{L|V,D}(l) = \varphi\big(h_{L|V,D}(V, D, l)\big) \left|\frac{\partial h_{L|V,D}(V, D, l)}{\partial l}\right|,
\label{eq:norm_flow_L_density}
\end{align}
where $\varphi(\cdot)$ denotes the standard normal density function and $\left|\frac{\partial h_{L|V,D}(V, D, l)}{\partial l}\right|$ is the absolute value of the derivative of the normalizing transformation with respect to $l$. This derivative captures the extent to which the transformation modifies the target variable so that it conforms to a standard normal distribution.

To illustrate further, suppose that the mediator $L$ is conditionally normally distributed with conditional mean $\mu_{L|V,D} = \beta_0+\beta_1V+\beta_2D$ and constant conditional variance $\sigma^2$. In this case, a normalizing flow for $L$, given $V$ and $D$, is a simple linear transformation:
\begin{align}
Z_L &= h_{L|V,D}(V, D, L; \beta,\sigma) \nonumber \\
    &= \frac{L - (\beta_0+\beta_1V+\beta_2D)}{\sigma}.
\label{eq:linear_norm_flow_for_L}
\end{align}
The inverse of this transformation is given by:
\begin{align}
L &= h_{L|V,D}^{-1}(V, D, Z_L; \beta, \sigma) \nonumber \\
  &= Z_L\sigma + (\beta_0+\beta_1V+\beta_2D).
\label{eq:linear_inv_norm_flow_for_L}
\end{align}
And the conditional distribution of $L$ can be expressed using the change-of-variables formula as follows:
\begin{align}
f_{L|V,D}(l) &= \varphi\bigl(h_{L|V,D}(v, d, l; \beta, \sigma)\bigr) \left|\frac{\partial h_{L|V,D}(v, d, l; \beta, \sigma)}{\partial l}\right| \nonumber \\
             &= \frac{1}{\sqrt{2\pi}} \exp\Bigl(-\frac{z_L^2}{2}\Bigr) \left|\frac{1}{\sigma}\right|,
\label{eq:linear_norm_flow_L_density}
\end{align}
where $\varphi(\cdot)$ is defined as before and $z_L = h_{L|V,D}(v, d, l; \beta, \sigma)$. Finally, to simulate values of $L$, we could draw Monte Carlo samples from the standard normal distribution, denoted by $\tilde{Z}_L$, and then apply the inverse transformation to obtain $\tilde{L} = \tilde{Z}_L\sigma + (\beta_0+\beta_1V+\beta_2D)$.

This example illustrates a situation where the conditional distribution of $L$ given $V$ and $D$ follows a known parametric form---specifically, $L \sim \mathcal{N}(\beta_0+\beta_1V+\beta_2D,\sigma^2)$. In practice, however, the true conditional distribution of $L$ and its dependence on $V$ and $D$ are typically unknown. By extension, the form of the normalizing transformation $h_{L|V,D}(V, D, L)$ is also unknown and potentially quite complex. To flexibly model this function without imposing strong parametric assumptions, we employ a UMNN (\citealt{Huang2018NAF, wehenkel2019UMNN}). UMNNs are both invertible and differentiable, and they are capable of accurately approximating any monotonic transformation of one variable to another (\citealt{Huang2018NAF, wehenkel2019UMNN, wehenkel2020GNF}).

The architecture of a UMNN is grounded in the principle that any monotonic function must have a strictly positive derivative, which can then be integrated to obtain the desired transformation. When applied to model the transformation of $L$ to a standard normal variable $Z_L$, conditional on $V$ and $D$, the network can be represented as follows:
\begin{align}
h_{L|V,D}(V, D, L; \tau, \omega) = \int_{0}^{L} \theta\left(t; \eta\left(V,D; \tau\right); \omega\right) \, \mathrm{d}t + \alpha\left(\eta\left(V,D; \tau\right)\right).  \label{seq:UMNN_for_L}
\end{align}
In this formulation, $\eta\left(V, D; \tau\right)$ is referred to as the \textit{embedding network}. It takes the conditioning variables $V$ and $D$ as inputs, transforms them using a set of weights, denoted by $\tau$, and then applies the rectified linear unit (ReLU) activation function to produce its output. Substantively, the embedding network captures how the distribution of the mediator $L$ varies as a function of the baseline confounders $V$ and the treatment $D$  (\citealt{balgi_deep_2025}).

The output from the embedding network is used to generate an offset term, denoted by $\alpha\left(\eta\left(V, D; \tau\right)\right)$, and it also serves as input to a second neural network, denoted by $\theta\left(t; \eta\left(V, D; \tau\right); \omega\right)$. Referred to as the \textit{integrand network}, $\theta\left(t; \eta\left(V, D; \tau\right); \omega\right)$ models the derivative, at point $t$, of a monotonic function designed to transform $L$ into a standard normal variable, given the input from the embedding network. To this end, the network transforms its input using a different set of weights, denoted by $\omega$, and an activation function known as the exponential linear unit incremented by one (ELUPlus), which generates an output that is strictly positive. Numerically integrating this output from $t=0$ to the observed value of $L$, and then adding the offset term, yields the transformed variable $Z_L$  (\citealt{balgi_deep_2025}).\footnote{Specifically, this integration is performed using Clenshaw-Curtis quadrature (\citealt{clenshaw1960method}).}

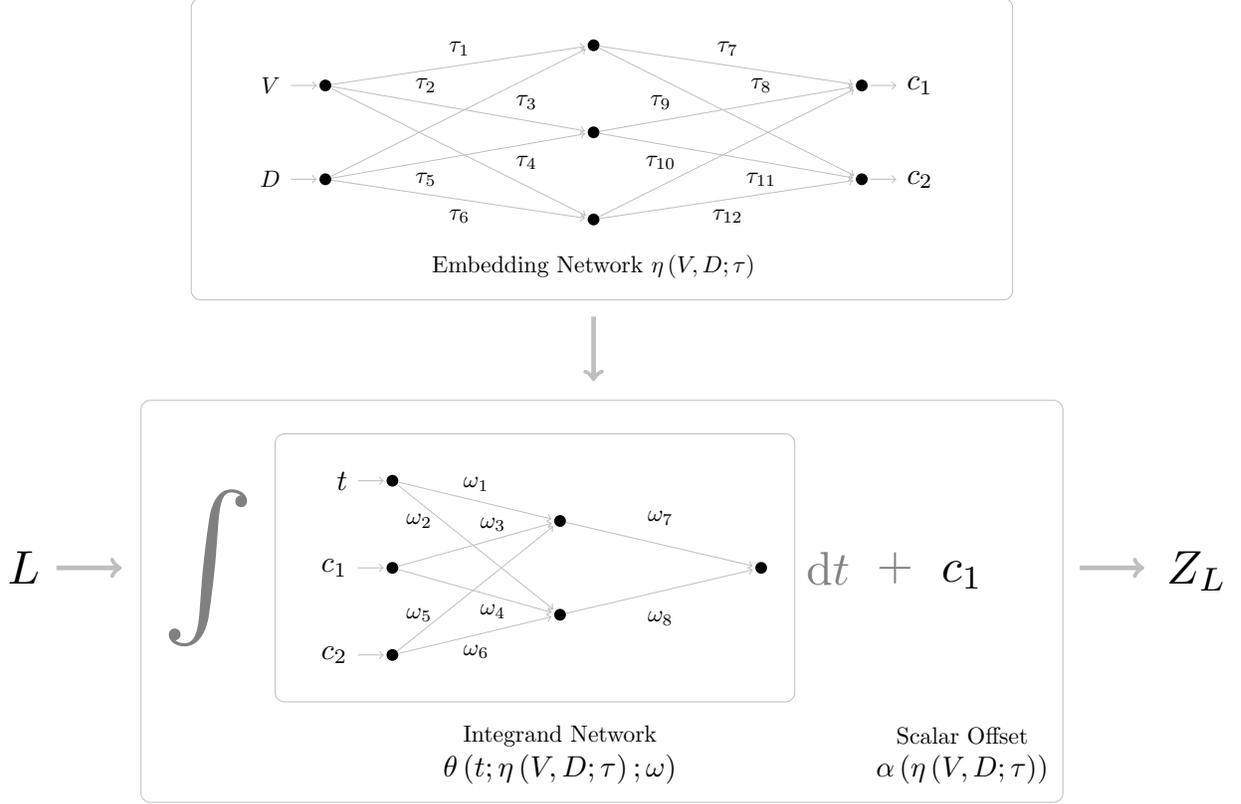
\begin{figure}[!t]
\begin{centering}
\resizebox{1.02\textwidth}{!}{
    \begin{tikzpicture}[shorten >=1pt,->,draw=gray!50]
        
        \tikzstyle{every pin edge}=[<-,shorten <=1pt]
        \tikzstyle{neuron}=[circle,fill=black,minimum size=5pt,inner sep=0pt]
        \tikzstyle{input neuron}=[neuron, fill=black];
        \tikzstyle{output neuron}=[neuron, fill=black];
        \tikzstyle{hidden neuron}=[neuron, fill=black];
        \tikzstyle{annot} = [text width=4em, text centered]
    
        \begin{scope}
            \foreach \name / \y in {V/-0.7,D/0.7}
            \path[yshift=0cm]
            node[input neuron, pin=left:$\name$] (I-\name) at (0,-\y cm) {};
        
            \foreach \name / \y in {1/-1.3,2/0,3/1.3}
            \path[yshift=0cm, xshift=1.5cm]
            node[hidden neuron] (H-\name) at (2.5cm,-\y cm) {};
        
            \node[output neuron, pin={[pin edge={->}]right:\scalebox{1.25}{$c_{1}$}}, right of=H-2] (O1) at (7cm, 0.7cm) {};
            \node[output neuron, pin={[pin edge={->}]right:\scalebox{1.25}{$c_{2}$}}, right of=H-2] (O2) at (7cm, -0.7cm) {};
        
            \foreach \source in {V,D}
                \foreach \dest in {1,2,3}
                    \path (I-\source) edge (H-\dest);
        
            \foreach \source in {1,2,3}
                \path (H-\source) edge (O1);
            \foreach \source in {1,2,3}
                \path (H-\source) edge (O2);
    
            \draw[rounded corners, solid] (-2cm, -2.5cm) rectangle (10.25cm, 2cm);
    
            \node[scale=1] (psi1) at (2cm, 1.25cm) {$\tau_1$};
            \node[scale=1] (psi2) at (1.5cm, 0.7cm) {$\tau_2$};
            \node[scale=1] (psi3) at (3cm, 0.45cm) {$\tau_3$};
            \node[scale=1] (psi4) at (3cm, -0.45cm) {$\tau_4$};
            \node[scale=1] (psi5) at (1.5cm, -0.7cm) {$\tau_5$};
            \node[scale=1] (psi6) at (2cm, -1.25cm) {$\tau_6$};
            \node[scale=1] (psi7) at (6cm, 1.25cm) {$\tau_7$};
            \node[scale=1] (psi8) at (6.5cm, 0.7cm) {$\tau_8$};
            \node[scale=1] (psi9) at (5cm, 0.45cm) {$\tau_{9}$};
            \node[scale=1] (psi10) at (5cm, -0.45cm) {$\tau_{10}$};
            \node[scale=1] (psi11) at (6.5cm, -0.7cm) {$\tau_{11}$};
            \node[scale=1] (psi12) at (6cm, -1.25cm) {$\tau_{12}$};
    
            \node at (4cm, -2cm) {Embedding Network $\eta\left(V,D;\tau\right)$};
            
        \end{scope}
    
        \begin{scope}[xshift=1cm, yshift=-6.5cm]
            \node[input neuron, pin=left:\scalebox{1.25}{$t$}] (I-1) at (0, 1.3cm) {};
            \node[input neuron, pin=left:\scalebox{1.25}{$c_{1}$}] (I-2) at (0, 0cm) {};
            \node[input neuron, pin=left:\scalebox{1.25}{$c_{2}$}] (I-3) at (0, -1.3cm) {};
            
            \foreach \name / \y in {1/-0.7,2/0.7}
            \path[yshift=0cm, xshift=2.5cm]
            node[hidden neuron] (H-\name) at (0cm,-\y cm) {};
        
            \node[output neuron, right of=H-1] (O) at (4.5cm, 0cm) {};
        
            \foreach \source in {1,2,3}
                \foreach \dest in {1,2}
                    \path (I-\source) edge (H-\dest);
        
            \foreach \source in {1,2}
                \path (H-\source) edge (O);
    
            \draw[rounded corners, solid] (-1.75cm, -2cm) rectangle (6cm, 2cm);
            \draw[rounded corners, solid] (-3.75cm, -3.5cm) rectangle (10cm, 2.5cm);
                
            \node at (2.5cm, -2.5cm) {Integrand Network};
            \node[scale=1.25] (h3) at (2.5cm, -3cm) {$\theta\left(t;\eta\left(V,D;\tau\right);\omega\right)$};\        
            \node at (8.5cm, -2.5cm) {Scalar Offset};
            \node[scale=1.22] at (8.5cm, -3cm) {$\alpha\left(\eta\left(V,D;\tau\right)\right)$};
        
            \node[scale=6, color=gray] at (-2.75cm, 0cm) {$\int$};
            \node[scale=2, color=gray] at (6.5cm, 0cm) {d$t$};
            \node[scale=2, color=gray] at (7.5cm, 0cm) {$+$};
            \node[scale=2, color=black] at (8.5cm, -0.1cm) {$c_{1}$};
    
            \node[scale=1] (phi1) at (1.25cm, 1.25cm) {$\omega_1$};
            \node[scale=1] (phi2) at (0.4cm, 0.7cm) {$\omega_2$};
            \node[scale=1] (phi4) at (1.5cm, 0.65cm) {$\omega_3$};
            \node[scale=1] (phi3) at (1.5cm, -0.65cm) {$\omega_4$};
            \node[scale=1] (phi5) at (0.4cm, -0.7cm) {$\omega_5$};
            \node[scale=1] (phi6) at (1.25cm, -1.25cm) {$\omega_6$};
            \node[scale=1] (phi7) at (4cm, 0.75cm) {$\omega_7$};
            \node[scale=1] (phi8) at (4cm, -0.75cm) {$\omega_8$};
            
            \node[scale=2] (V3) at (-5.5cm, 0cm) {$L$};
            \draw[->, line width=2pt] (V3) -- (-4cm, 0cm);
            \node[scale=2] (Z3) at (12cm, -0.05cm) {$Z_L$};
            \draw[->, line width=2pt] (10.25cm, 0cm) -- (11.25cm, 0cm);
        \end{scope}
    
        \draw[->, line width=2pt] (4cm, -2.75cm) -- (4cm, -3.75cm);
    
    \end{tikzpicture}
}
\caption{An Unconstrained Monotonic Neural Network (UMNN) for the Mediator $L$ Depicted using a Computational Graph. \label{fig:UMNN_comp_graph}}
\medskip{}
\par\end{centering}
Note: This figure illustrates a simple UMNN for the mediator $L$. In this model, the embedding network consists of an input layer with two nodes, corresponding to $V$ and $D$. This is followed by a single hidden layer with three nodes and an output layer with two nodes, labeled $c_1$ and $c_2$. The outputs of the embedding network serve as inputs to the integrand network. Specifically, the integrand network has an input layer with three nodes: $c_1$ and $c_2$ from the embedding network, along with a set of integration points $t$. This input layer is followed by a single hidden layer with two nodes and an output layer with a single node. The output of the integrand network is integrated from $t=0$ to $L$ using Clenshaw-Curtis quadrature, which yields the transformed variable $Z_L$. The parameters $\{\tau, \omega\}$ represent the weights associated with the embedding and integrand networks.
\end{figure}

To summarize, the expression in Equation \ref{seq:UMNN_for_L} consists of two neural networks working in tandem. The embedding network models how the distribution of $L$ varies as a function of $V$ and $D$, while the integrand network adjusts its shape to match the standard normal distribution. Figure \ref{fig:UMNN_comp_graph} illustrates an example of this UMNN with a relatively simple architecture in the form of a computational graph.

Although the embedding and integrand networks shown in Figure \ref{fig:UMNN_comp_graph} each contain only a single hidden layer with just a few nodes, more complex architectures can be constructed using multiple hidden layers, varying numbers of nodes per layer, and different patterns of connectivity between layers. Architectures with additional layers, nodes, and connections are generally more expressive---that is, better able to approximate a wide range of transformations. When configured with a sufficiently expressive architecture, UMNNs serve as universal density approximators, meaning they can accurately model any continuous distribution regardless of its complexity \citep{Huang2018NAF, wehenkel2019UMNN}. Following Balgi et al. \citeyearpar{balgi_deep_2025}, we recommend architectures for the embedding and integrand networks that include at least four hidden layers, a minimum of 10 to 20 nodes per layer, and full connectivity between layers.

To train a UMNN, the weights in the embedding and integrand networks are randomly initialized and then iteratively updated to minimize a loss function. The loss function used for training is the negative log-likelihood, derived from the change-of-variables formula. Substantively, this loss function quantifies how well the model transforms $L$ into a new variable $Z_L$ that follows a standard normal distribution. Formally, it can be expressed as follows:
\begin{align}
-\text{log}\mathcal{L}\left(\tau,\omega\right) &= -\ln\left(\prod_{i=1}^{n}\varphi\bigl(h_{L|V,D}(v_i, d_i, l_i; \tau, \omega)\bigr) \left|\frac{\partial h_{L|V,D}(v_i, d_i, l_i; \tau, \omega)}{\partial l_i}\right|\right) \nonumber \\
&= -\sum_{i=1}^{n} \ln \left( \varphi\bigl(h_{L|V,D}(v_i, d_i, l_i; \tau, \omega)\bigr) \right) -\sum_{i=1}^{n} \ln \left( \left|\frac{\partial h_{L|V,D}(v_i, d_i, l_i; \tau, \omega)}{\partial l_i}\right| \right),
\label{seq:neg_LL_loss}
\end{align}
where $i = 1, \dots, n$ indexes the observations and $\varphi(\cdot)$ denotes the standard normal density function. The network weights $\{\tau,\omega\}$ are optimized using mini-batch stochastic gradient descent (SGD), an algorithm that iteratively updates the weights based on the gradient of the loss function. At each iteration, the algorithm processes a random subset of the training data and adjusts the weights incrementally to reduce the loss by a small amount. This process continues until further adjustments no longer reduce the loss, as evaluated on a separate validation sample held out from training \citep{lecun2015deep, chollet2021deep, goodfellow2016deep}.

After training the UMNN, it can then be used to generate simulated values from the conditional distribution of the mediator $L$, given the baseline confounders $V$ and the treatment $D$. This is accomplished by first drawing Monte Carlo samples from the standard normal distribution and then transforming them using the inverse of the trained UMNN. The inverse is computed using a bisection algorithm, which iteratively narrows the range of possible solutions until it identifies the inverse. For example, to simulate $J$ values from the conditional distribution of $L$, we first generate standard normal samples, denoted by $\tilde{Z}_{L,j}$ for $j=1, \dots, J$, and then transform them as follows:
\begin{align}
\tilde{L}_j &= h_{L|V,D}^{-1} \left( V, D, \tilde{Z}_{L,j}; \hat{\tau}, \hat{\omega} \right), \label{eq:MC_sample_L}
\end{align}
where the hats indicate that the network weights have been optimized via SGD. In this expression, $h_{L|V,D}^{-1}$ denotes the inverse of the normalizing transformation, and $V$ and $D$ represent observed values of the confounders and treatment, respectively. Applying $h_{L|V,D}^{-1}$ to $\tilde{Z}_{L,j}$ produces $\tilde{L}_j$, a Monte Carlo sample from the conditional distribution of $L$.

To simulate potential values of the mediator under a specific treatment condition, we follow the same sampling procedure as outlined above but set the exposure $D$ to a fixed value for everyone. For example, to generate simulated values for $L(d_1)$, we replace each sample member's treatment with $d_1$, while keeping their baseline confounders $V$ unchanged. We then transform Monte Carlo draws from the standard normal distribution using the following expression:
\begin{align}
\tilde{L}_{j}(d_1) = h_{L|V,D}^{-1} \left( V, d_1, \tilde{Z}_{L,j}; \hat{\tau}, \hat{\omega} \right) \quad \text{for} \: j=1, \dots, J, \label{eq:MC_sample_of_d1}
\end{align}
which yields $J$ simulated copies of the potential value for $L$ under treatment $d_1$.

Expanding this approach to estimate interventional, path-specific, and multivariate natural effects, we train normalizing flows not only for $L$ but also for $X$ and $Y$. That is, in the first step the estimation procedures outlined in Section \ref{sec:Parametric-Estimation}, we replace parametric models with normalizing flows for the conditional distributions of $L$, $X$, and $Y$. These normalizing flows can be trained simultaneously rather than separately by minimizing a negative log-likelihood based on the joint distribution of the data \citep{balgi_deep_2025, wehenkel2020GNF}. Once trained, the flows are used to generate simulated values for $L$, $X$, and $Y$ under different conditions, following the subsequent steps of the estimation procedures in Section \ref{sec:Parametric-Estimation} exactly. These values are then averaged together and compared to estimate the effects of interest. Thus, the simulation approach to estimating interventional, path-specific, and multivariate natural effects remains essentially unchanged, except that normalizing flows replace parametric models for the conditional distributions of the mediators and outcome in the initial step.

When estimating these effects using normalizing flows instead of parametric models, one additional modification to the estimation procedure is advisable. In the procedures outlined in Section \ref{sec:Parametric-Estimation}, $J$ copies of the relevant potential outcomes are simulated for each sample member. Computing $J \times n$ simulated values by inverting the normalizing flow with a bisection algorithm can be computationally expensive, especially for $J \geq 10^3$ and $n \gg 10^3$. A more efficient alternative is to draw a bootstrap sample (with replacement) of $b \ll J \times n$ cases from the sample data and generate a single simulated value per bootstrap observation. This adjustment reduces the computational cost of simulating values from the normalizing flows, particularly in large-sample applications. As long as $b$ is still sufficiently large (e.g., $b \geq 10^5$), this modification should not appreciably influence the precision of the resulting estimates.

In summary, the simulation approach to mediation analysis, when implemented using normalizing flows and UMNNs, is incredibly flexible. It can accommodate virtually any set of causal relationships among the treatment, mediators, and outcome, as well as virtually any type of distribution for these variables, all without imposing restrictive parametric assumptions. This flexibility, however, comes at a cost: UMNNs require large amounts of data for training and are susceptible to finite-sample bias. Moreover, the rate and conditions under which their effect estimates converge to their target estimands remain unknown. As a result, they are best suited for large-sample applications, where existing evidence suggests that their estimates are quite accurate with sample sizes of $n \geq 16,000$ across a range of data-generating processes. In these settings, the nonparametric bootstrap also appears to yield valid, or potentially conservative, confidence intervals. By contrast, in smaller samples or in applications where the available data are relatively sparse, their performance can degrade, resulting in biased estimates and invalid inferences (\citealt{balgi_deep_2025}).

\section{The Effect of Media Framing on Immigration Attitudes}

To illustrate the simulation approach using parametric models, we reanalyze data from Brader et al. (\citeyear{brader2008triggers}), building on related analyses by Imai et al. (\citeyear{imai2013identification}), Zhou and Yamamoto (\citeyear{zhou2019regression}), and \citeauthor{wodtke20XXmediation} Forthcoming). The original study investigated how negative media framing influences attitudes toward immigration in the United States. Drawing on a nationally representative sample of $n=354$ white, non-Hispanic adults, the researchers conducted a survey experiment in which participants were randomly assigned to read a mock news article about immigration. The experiment employed a $2\times2$ design, varying both the ethnicity of the immigrant featured in the article (white European versus Latino) and the framing of the story (emphasizing either the benefits or the costs of immigration). After reading the article, participants reported their beliefs about the harms of immigration, their level of anxiety about immigration, and their overall support for immigration.

Brader et al. (\citeyear{brader2008triggers}) found that exposure to a story featuring both a Latino immigrant and a negatively framed message emphasizing costs significantly reduced support for immigration. They further suggested that participant anxiety mediated this effect, whereas beliefs about the harms of immigration appeared to play a more limited explanatory role. However, their mediation analysis treated anxiety and beliefs about harm as causally independent, even though perceptions of harm likely contribute to increased anxiety (\citealt{imai2013identification, zhou2019regression}). In addition, their analysis relied on additive linear regression models applied to an outcome measured on an ordinal rather than cardinal scale, which can yield misleading results.

In contrast, our reanalysis allows for causal dependence between the two mediators and employs models better suited for ordinal data. Specifically, we apply the simulation approach to estimate interventional, multivariate natural, and path-specific effects of negative framing, treating perceptions of harm and anxiety as causally ordered mediators. We implement the simulation approach using parametric models rather than normalizing flows and UMNNs owing to the limited sample size, which would render these more flexible approaches unreliable. In this analysis, the outcome $Y$ denotes support for immigration, measured on a five-point scale; the treatment $D$ indicates exposure to a negatively framed article featuring a Latino immigrant; the first mediator $L$ measures beliefs about the harms of immigration on a seven-point scale; the second mediator $X$ captures participant anxiety on a ten-point scale; and $V$ represents a vector of baseline covariates, including gender, age, education, and income. The causal relationships we hypothesize between these variables align with those depicted earlier in the graphical model from Figure \ref{fig:simple_med_model}.

We first estimate the effects of interest using ordinal logit models for $L$, $X$, and $Y$, each with an additive linear predictor. We also estimate these effects using ordinal logit models that include two-way interactions between the treatment $D$ and the mediators $\{L,X\}$, wherever appropriate. After fitting these models, we use them to simulate $J=2000$ potential values of the mediators and outcome for each sample member, following the steps outlined in Section \ref{sec:Parametric-Estimation}. Results from this analysis are presented in Table \ref{tab:effects_on_immgr_support}, where all estimates have been translated to effect sizes denominated in standard deviation (SD) units and confidence intervals are based on the nonparametric bootstrap. A complete set of replication files for these estimates is available at \url{https://github.com/JesseZhou-1/Causal-Mediation-Analysis-with-Multiple-Mediators}.

\begin{table}[ht!]
    \begin{centering}
    \caption{Estimates for Interventional, Path-specific, and Multivariate Natural Effects of Negative Media Framing on Support for Immigration.}
    \label{tab:effects_on_immgr_support}
    \renewcommand{\arraystretch}{1.3}
    \begin{tabular}{{@{}p{5cm}>{\raggedleft\arraybackslash}p{4cm}>{\raggedleft\arraybackslash}p{4cm}@{}}}
    \toprule
    \noalign{\vskip -1.5ex}
    & \multicolumn{2}{c}{\multirow{2}{*}{\parbox{8cm}{\centering\textbf{Point Estimates and 95\% CIs}}}} \\
    \noalign{\vskip -1ex}
    \multirow{-2}{*}{\parbox{5cm}{\vspace{8ex}\textbf{Estimand}\vspace{2ex}}} & & \\
    \cline{2-3}
    \noalign{\vspace{1ex}}
    & \parbox[t]{4cm}{
    \raggedleft
    \textbf{Without}\\
    \textbf{Interactions}
} & \(\bm{D \times \{L, X\}}\) \textbf{Interactions} \\
    \midrule
        & & \\
        \multicolumn{3}{l}{\textbf{Interventional Effects}} \\
        $\quad OE(1,0)$ &  $-.46$ [$-.70$, $-.21$] & $-.46$ [$-.69$, $-.20$]\\
        $\quad IDE(1,0)$ & $-.35$ [$-.58$, $-.10$] & $-.40$ [$-.64$, $-.10$]\\
        $\quad IIE(1,0)$ & $-.11$ [$-.21$, $-.03$] & $-.06$ [$-.23$, .06]\\
         & & \\
        \multicolumn{3}{l}{\textbf{Total Effect}} \\
        $\quad ATE(1,0)$ & $-.44$ [$-.68$, $-.20$] & $-.45$ [$-.68$, $-.21$]\\
         & & \\
        \multicolumn{3}{l}{\textbf{Multivariate Natural Effects}} \\
        $\quad MNDE(1,0)$ & $-.25$ [$-.46$, $-.02$] & $-.28$ [$-.48$, $-.03$]\\
        $\quad MNIE(1,0)$ & $-.20$ [$-.35$, $-.05$] & $-.17$ [$-.36$, $-.02$]\\
         & & \\
        \multicolumn{3}{l}{\textbf{Path-specific Effects}} \\        
        $\quad PSE_{D \rightarrow Y}(1,0)$ & $-.25$ [$-.46$, $-.02$] & $-.28$ [$-.48$, $-.03$]\\
        $\quad PSE_{D \rightarrow L \rightsquigarrow Y}(1,0)$ &  $-.13$ [$-.27$, .00] &  $-.13$ [$-.30$, .00]\\
        $\quad PSE_{D \rightarrow X \rightarrow Y}(1,0)$ &  $-.06$ [$-.13$, $-.01$] &  $-.04$ [$-.15$, .04]\\      
        \bottomrule
    \end{tabular}
\par\end{centering}
\medskip{}
Note: Estimates are expressed in standard deviation units. The numbers in parentheses represent 95\% confidence intervals, which were computed using the nonparametric bootstrap with 2000 replications.
\small
\end{table}

Replicating the central result of the original study, we find a large total effect of negative framing on support for immigration. Exposure to a story about a Latino immigrant that emphasizes the costs of immigration decreases support by roughly 0.5 SDs, on average. Our analysis of the multivariate natural and path-specific effects suggests that more than half of this treatment impact is not explained by either perceived harms or anxiety. For example, the estimated $PSE_{D\to Y}(1,0)$ based on models with exposure-mediator interactions is $-0.28$ SDs, which accounts for about 60\% of the total effect.

The independent mediating effect of anxiety, net of beliefs about the harms of immigration, is captured by the $PSE_{D\to X \to Y}(1,0)$. According to the models with interactions, this effect is relatively modest---specifically, the estimated effect through anxiety is $-0.04$ SDs, representing only about 10\% of the total effect. By contrast, the estimated $PSE_{D\to X \rightsquigarrow Y}(1,0)$, which captures mediation via perceptions of harm, is larger across all specifications. 

Taken together, these findings raise questions about the conclusion that anxiety is the primary mediator of negative framing effects on support for immigration. When perceptions of harm are properly incorporated into an analysis using a more defensible modeling strategy, the unique explanatory role of anxiety appears comparatively limited, while mediation through perceived harm is more pronounced.

\section{The Effect of Prenatal Care on the Risk of Preterm Birth}

To illustrate the simulation approach using normalizing flows and UMNNs, we revisit an analysis by VanderWeele et al. \citeyearpar{vanderweele2014effectdecomp} evaluating the impact of prenatal care on the risk of preterm birth. Their analysis examined whether maternal smoking and pre-eclampsia mediate the effect of receiving adequate prenatal care, using an IPW approach to estimate interventional and path-specific effects. Because IPW can be challenging to implement with many-valued mediators, they treated both mediators as binary variables, despite the availability of more detailed data on maternal smoking, measured as the number of cigarettes smoked daily during pregnancy. In our reanalysis, we incorporate this additional information to better capture variation in smoking behavior, leveraging the flexibility of the simulation approach to seamlessly accommodate many different types of variables.

The original study analyzed a sample of $n=2,629,247$ newborns and their mothers, based on birth certificate data from 2003. The large sample size makes these data well-suited for normalizing flows and UMNNs, which can approximate complex distributions very accurately with this much data. The key variables consist in a set of baseline confounders $V$, including measures of maternal age, marital status, educational attainment, and ethnicity; a binary treatment $D$, coded 1 if the mother received adequate prenatal care and 0 if the care was inadequate, as determined by a modified version of the Adequacy of Prenatal Care Utilization Index (\citealt{kotelchuck1994}); an initial mediator $L$, which measures the daily number of cigarettes smoked during pregnancy on an ordinal scale, coded 0 for non-smokers, 1 for those who smoked 1-5 cigarettes, 2 for 6-10 cigarettes, 3 for 11-20 cigarettes, 4 for 21-30 cigarettes, and 5 for $30+$ cigarettes per day; a second mediator $X$, coded 1 if the mother was diagnosed with pre-eclampsia and 0 otherwise; and finally, an outcome variable $Y$, coded 1 for a preterm birth and 0 otherwise.

The hypothesized causal relationships among these variables align with the graphical model presented earlier in Figure \ref{fig:simple_med_model}. In this model, the baseline confounders ($V$) influence all downstream variables. The adequacy of prenatal care ($D$) is hypothesized to affect both maternal smoking ($L$) and pre-eclampsia ($X$), which in turn influence the risk of preterm birth ($Y$). Additionally, maternal smoking ($L$) is hypothesized to have an effect on pre-eclampsia ($X$), making it both a mediator and a treatment-induced confounder of the $X \rightarrow Y$ relationship. Prenatal care may also influence the risk of preterm birth directly.

To estimate the interventional, path-specific, and multivariate natural effects of prenatal care on the risk of preterm birth, we implement the simulation approach using normalizing flows and UMNNs. Our model architecture consists of an embedding network with five hidden layers, containing 100, 90, 80, 70, and 60 nodes, respectively. The integrand network follows a similar structure, with five hidden layers consisting of 60, 50, 40, 30, and 20 nodes each. We use the same architecture for both mediators and the outcome, training all these flows simultaneously via SGD. To optimize the network weights, we allocate 80\% of the data for training and use the remaining 20\% to compute the validation loss and determine when to terminate the SGD algorithm. After training, we simulate potential values for the mediators and outcome following the steps outlined in Section \ref{sec:Parametric-Estimation}, with one modification: to reduce computational costs, we draw a bootstrap sample of $b=10^5$ observations from the original data and then generate a single simulated value for each variable of interest per observation.

Because the loss function for deep neural networks is typically not convex, the training algorithm may converge to a local rather than global minimum when optimizing the weights. To mitigate this problem, we train five models with identical architectures but different random initializations for the network weights. The resulting estimates were very similar across models, so we report only one set of results from the model with the lowest validation loss.

For comparison, we also compute estimates using the parametric simulation approach. Specifically, we model maternal smoking ($L$) using ordinal logistic regression, while pre-eclampsia ($X$) and preterm birth ($Y$) are both modeled using binary logistic regression. Each of these models is specified with all two-way interactions among the treatment, mediators, and covariates. After fitting them by maximum likelihood, we use these models to simulate $J=2000$ potential values of the mediators and outcome for each observation, following the steps outlined in Section \ref{sec:Parametric-Estimation} exactly. Given the large sample size, we do not report inferential statistics in this analysis, as the degree of sampling error is negligible. Instead, we focus on effect sizes and their practical significance. Replication files are available at \url{https://github.com/JesseZhou-1/Causal-Mediation-Analysis-with-Multiple-Mediators}.

Results from this analysis are presented in Table \ref{tab:effects_on_preterm_birth}. The first column reports estimates derived from normalizing flows modeled with UMNNs, while the second column presents estimates based on parametric models for the mediators and outcome. The similarity between these two sets of estimates suggests that the results are highly robust to potential violations of the functional form assumptions imposed by the parametric models.

\begin{table}[ht!]
    \begin{centering}
    \caption{Estimates for Interventional, Path-specific, and Multivariate Natural Effects of Adequate versus Inadequate Prenatal Care on Preterm Birth.}
    \label{tab:effects_on_preterm_birth}
    \renewcommand{\arraystretch}{1.3}
    \begin{tabular}{@{}p{3.25cm}>{\raggedleft\arraybackslash}p{4cm}>{\raggedleft\arraybackslash}p{4cm}@{}}
        \toprule
        \textbf{Estimand} & \textbf{UMNNs} & \textbf{Parametric Models} \\
        \midrule
        & & \\
        \multicolumn{3}{l}{\textbf{Interventional Effects}} \\
        $\quad OE(1,0)$ & $-.061$ & $-.060$ \\
        $\quad IDE(1,0)$ & $-.060$ & $-.060$ \\
        $\quad IIE(1,0)$ &  .000 &  .000 \\
         & & \\
        \multicolumn{3}{l}{\textbf{Total Effect}} \\
        $\quad ATE(1,0)$ & $-.061$ & $-.060$ \\
         & & \\
        \multicolumn{3}{l}{\textbf{Multivariate Natural Effects}} \\
        $\quad MNDE(1,0)$ & $-.059$ & $-.059$ \\
        $\quad MNIE(1,0)$ & $-.001$ & $-.001$ \\
         & & \\
        \multicolumn{3}{l}{\textbf{Path-specific Effects}} \\        
        $\quad PSE_{D \rightarrow Y}(1,0)$ & $-.059$ & $-.059$ \\
        $\quad PSE_{D \rightarrow L \rightsquigarrow Y}(1,0)$ & $-.001$ & $-.001$ \\
        $\quad PSE_{D \rightarrow X \rightarrow Y}(1,0)$ & .000 & .000 \\      
        \bottomrule
    \end{tabular}
\par\end{centering}
\medskip{}
Note: The UMNN estimates are based on the model with the lowest validation loss. The parametric estimates are based on an ordinal logit model for maternal smoking, and binary logit models for pre-eclampsia and preterm birth.
\small
\end{table}

These findings are also broadly consistent with those originally reported by VanderWeele et al. \citeyearpar{vanderweele2014effectdecomp}. The estimated overall effect, $\widehat{OE}(1,0)$, indicates that receiving adequate prenatal care reduces the risk of preterm birth by approximately 6 percentage points. The estimated interventional direct effect, $\widehat{IDE}(1,0)$, is nearly identical, while the interventional indirect effect, $\widehat{IIE}(1,0)$, is negligible, suggesting that pre-eclampsia plays little to no mediating role in this relationship.

The total, multivariate, and path-specific effects follow a similar pattern. The estimated total effect, $\widehat{ATE}(1,0)$, suggests that adequate prenatal care reduces the risk of preterm birth by roughly 6 percentage points. The multivariate natural direct effect, $\widehat{MNDE}(1,0)$, which corresponds with the path-specific effect from treatment directly to the outcome, $\widehat{PSE}_{D \rightarrow Y}(1,0)$, is nearly identical. In contrast, both the multivariate natural indirect effect, $\widehat{MNIE}(1,0)$, and the path-specific effects operating through maternal smoking and pre-eclampsia---$\widehat{PSE}_{D \rightarrow L \rightsquigarrow Y}(1,0)$ and $\widehat{PSE}_{D \rightarrow X \rightarrow Y}(1,0)$, respectively---are approximately equal to zero. Taken together, these results suggest that the effect of prenatal care on the risk of preterm birth is not mediated by maternal smoking or pre-eclampsia but instead operates through other, unmeasured mechanisms.

\begin{figure}[t!]
  \begin{centering}
\includegraphics[width=\linewidth,height=\linewidth,keepaspectratio]{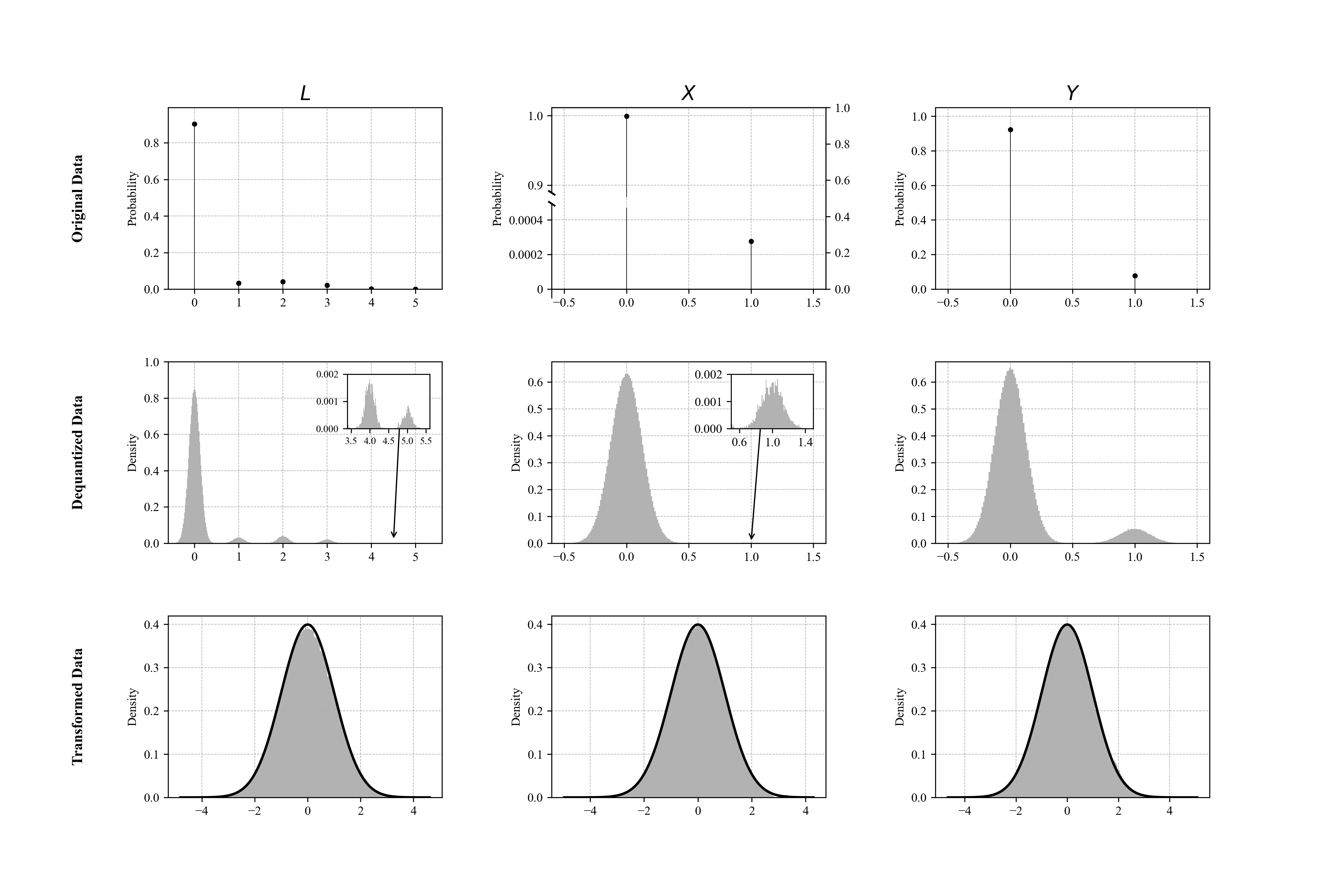}
\caption{Dequantization and UMNN Transformation of the Mediators and Outcome}
\label{fig:trasform_plot}
\medskip{}
\par\end{centering}
Note: All plots are based on the model with the lowest validation loss. The black line in the third row represents the probability density of the standard normal distribution. $L$ is an ordinal variable that encodes the frequency of maternal smoking, $X$ is a binary variable indicating whether a mother was diagnosed with pre-eclampsia, and $Y$ is a binary variable indicating a preterm birth. 
\end{figure}

Figure \ref{fig:trasform_plot} presents a series of plots illustrating how the trained normalizing flows transform the original data. The first row shows the empirical distributions of the two mediators and the outcome. The second row displays histograms for these variables after they have been dequantized, which involves introducing a small amount of normally distributed noise to their original integer values. The final row presents a set of histograms for the transformed variables after the dequantized data have been mapped through the trained UMNNs. These plots serve as a graphical diagnostic of model performance. In the third row of the figure, the transformed data closely follow the standard normal distribution, represented by the solid black line, which demonstrates that our flows perform extremely well in this application.

To summarize, our reanalysis shows that both the parametric and nonparametric implementations of the simulation approach yield similar estimates for the effects of prenatal care on preterm birth. While the parametric approach is flexible and computationally efficient, it relies on specific functional form assumptions that may not always hold. In contrast, the nonparametric approach based on normalizing flows and UMNNs provides a data-driven alternative capable of modeling complex relationships and distributions without imposing restrictive assumptions on the data. The close alignment between the two methods suggests that the estimated effects are robust to different modeling choices in this case. In general, our findings highlight the promise of normalizing flows and UMNNs for analyzing causal mediation, particularly in large-sample applications.

\section{Discussion}

In this study, we introduced a simulation approach for causal mediation analysis with multiple mediators, allowing for estimation of multivariate natural, interventional, and path-specific effects. Our method extends the simulation estimator developed by Imai and colleagues \citeyearpar{imai2010general, imai2011unpacking} for analyses of univariate natural effects to applications with multiple mediators, including those that are causally related. To implement this approach, we outlined two strategies. The first relies on parametric models for each mediator and the outcome. The second approach trains a special class of deep neural networks to learn an invertible normalizing transformation of the data, which is then used to simulate the potential outcomes of interest. Both methods are highly flexible and can accommodate treatments, mediators, and outcomes that are either continuous or discrete (i.e., binary, ordinal, polytomous, and so on).

We illustrated the proposed methods through two applications: a reanalysis of whether the effects of negative media framing on attitudes toward immigration are mediated by anxiety and perceptions of harm (\citealt{brader2008triggers}), and an evaluation of whether the effects of prenatal care on preterm birth are mediated by maternal smoking and pre-eclampsia (\citealt{vanderweele2014effectdecomp}). These examples demonstrated the versatility of our approach as a highly general and adaptable method for mediation analysis with multiple mediators. And although we focused on applications with only two mediators for clarity and illustration, both the parametric and nonparametric implementations can be easily extended for more complex settings with additional mediators.

While the simulation approach to analyzing mediation offers many advantages, it is also subject to several limitations. The first is its reliance on strong identification assumptions. Depending on the target estimands, the simulation approach variously requires that all relevant confounders affecting the treatment-outcome, treatment-mediator, mediator-mediator, and/or the mediator-outcome relationships are observed and properly accounted for. In observational studies, where neither the treatment nor the mediators are experimentally manipulated, these assumptions can be especially onerous. Any violation of them may lead to biased estimates, whether using parametric models or UMNNs. Even in experimental studies, where treatment is randomized and the assumptions of no unobserved treatment-outcome and treatment-mediator confounding are satisfied by design, the possibility that the mediator-mediator or mediator-outcome relationships are confounded still remains. Thus, regardless of the study design, researchers must critically assess the plausibility of these assumptions when applying the simulation approach.

When key identification assumptions are met, the choice between implementing the simulation approach with parametric models versus normalizing flows and UMNNs also involves trade-offs. Parametric models provide a straightforward means of estimating and drawing inferences about mediation effects, but they rely on strong functional form assumptions. If these modeling assumptions are violated, the resulting estimates may be biased and inconsistent. By contrast, normalizing flows and UMNNs avoid restrictive parametric assumptions, and they provide greater flexibility for modeling complex relationships and distributions. However, this flexibility comes at a cost: their estimates are susceptible to finite-sample bias. Monte Carlo experiments suggest that normalizing flows and UMNNs yield reliable estimates with sample sizes in the tens of thousands but may suffer from nontrivial bias when applied to data with only a few thousand observations (\citealt{balgi_deep_2025}). Consequently, researchers must carefully balance the risk of model misspecification with the parametric approach against the data requirements of UMNNs.

Another limitation of normalizing flows and UMNNs is that the conditions and rate at which their estimates converge to the target estimands of interest remain largely unknown. In theory, these networks can approximate any distribution to an arbitrary degree of accuracy, given a sufficiently expressive architecture and enough data for training (\citealt{Huang2018NAF, wehenkel2019UMNN}), but the specific conditions under which UMNNs yield consistent estimates of causal effects have yet to be formally established. Monte Carlo evidence suggests that their bias and variance steadily decrease with the sample size, albeit more slowly than for traditional parametric estimators, and that bootstrap confidence intervals can provide accurate coverage in large samples (\citealt{balgi_deep_2025}). However, the absence of theoretical guarantees regarding the shape, central tendency, and dispersion of the sampling distribution limits the utility of UMNNs for statistical inference. Until valid inferential procedures are firmly established, this approach is best suited for applications where the sample size is large enough to obviate concerns about sampling error.

The limitations outlined above suggest several promising directions for future research. For example, it would be valuable to systematically compare the performance of our simulation approach with alternative estimation techniques, such as inverse probability weighting (\citealt{vanderweele2014effectdecomp}), regression imputation (\citealt{zhou2023tracing}), and linear modeling (\citealt{hayes2017introduction, vanderweele2015explanation, wodtke2020effect}). Such comparisons could clarify the conditions under which each method performs best in terms of bias and efficiency. These analyses could also evaluate the robustness of parametric estimates to various forms of model misspecification and investigate how sensitive UMNN estimates are to the choice of different architectures during training.

In addition, future research should attempt to formally establish the asymptotic properties of effect estimates derived from normalizing flows and UMNNs. Although existing Monte Carlo evidence is promising, clearly identifying the conditions under which these estimates are consistent and thoroughly characterizing their sampling distribution would significantly enhance the utility of UMNNs for causal inference in practice (\citealt{balgi_deep_2025}). Such theoretical developments would facilitate more defensible inferences across a much broader range of empirical applications.

Finally, incorporating sensitivity analyses that account for different forms of unobserved confounding would further strengthen the robustness of the simulation approach with multiple mediators. Although bias formulas for multivariate natural, interventional, and path-specific effects have already been derived, their application in sensitivity analyses often relies on restrictive assumptions about the structure and form of unobserved confounding in order to reduce complexity (\citeauthor{wodtke20XXmediation} Forthcoming). A more flexible approach to sensitivity analysis is available when estimating causal effects from normalizing flows and UMNNs (\citealt{balgi_deep_2025}), but adapting it for use with parametric models is not straightforward. Future research should explore strategies for generalizing and integrating different methods of sensitivity analysis with the simulation approach to analyzing mediation with multiple mediators.

In conclusion, the simulation estimators outlined in this study are highly versatile and offer considerable promise for improving analyses of causal mediation, particularly in settings with multiple mediators. The parametric approach can accommodate many different types of treatments, mediators, and outcomes, as well as complex relationships among these variables. Moreover, it can yield accurate effect estimates across a range of applications, provided that the requisite models are correctly specified. In contrast, the approach based on normalizing flows and UMNNs avoids restrictive parametric assumptions and largely obviates concerns about model specification, but it requires larger samples for accurate estimation. Together, by balancing their respective strengths and limitations, these methods represent a significant advance in causal mediation analysis.

\bibliographystyle{apalike}
\bibliography{multiMedSim.bib}

\end{document}